\documentclass[%
preprint,
 amsmath,amssymb,
 aps,
superscriptaddress,
showkeys,
superscriptaddress
]{revtex4-1}

\usepackage{graphicx,textcomp}
\usepackage{dcolumn}
\usepackage{bm}
\usepackage{bbold}
\usepackage{mwe}
\usepackage{color}
\usepackage{algorithm}
\usepackage{algorithmicx}
\usepackage{algpseudocode}
\usepackage{caption}
\usepackage{array}
\usepackage{rotating}
\usepackage[table]{xcolor}
\usepackage{hyperref}
\usepackage{changes}

\usepackage[utf8]{inputenc}
\usepackage[T1]{fontenc}
\usepackage{mathptmx}

\usepackage{academicons}

\newcommand{\orcid}[1]{\href{https://orcid.org/#1}{\textcolor[HTML]{A6CE39}{\aiOrcid}}}

\algnewcommand\INPUT{\item[\textbf{Input:}]}%
\algnewcommand\GIVEN{\item[\textbf{Given:}]}%
\algnewcommand\OUTPUT{\item[\textbf{Output:}]}%

\usepackage{changes}

\begin{document}

\title[A unified and automated approach to attractor reconstruction]{A unified and automated approach to attractor reconstruction}

\author{K.~H.~Kraemer}
\homepage[Visit: ]{https://www.pik-potsdam.de/members/hkraemer}
\email[E-mail me at: ]{hkraemer@pik-potsdam.de}
\affiliation{Potsdam Institute for Climate Impact Research, Member of the Leibniz Association, Telegraphenberg A 31
14473 Potsdam, EU
}
\affiliation{Institute of Physics and Astronomy, University of Potsdam, Karl-Liebknecht-Str.~24--25, 14476 Potsdam, EU}
\affiliation{Institute of Geosciences, University of Potsdam, Karl-Liebknecht-Str.~24--25, 14476 Potsdam, EU
}



\author{G.~Datseris}
\affiliation{Max Planck Institute for Meteorology, Bundesstr. 53, 20146 Hamburg, EU 
}

\author{J.~Kurths}

\affiliation{Potsdam Institute for Climate Impact Research, Member of the Leibniz Association, Telegraphenberg A 31
14473 Potsdam, EU
}
\affiliation{Institute of Physics, Humboldt Universit\"at zu Berlin, 12489 Berlin, EU
}
   
\author{I.~Z.~Kiss}
\affiliation{Department of Chemistry, Saint Louis University, St. Louis, USA
}

\author{J.~L.~Ocampo-Espindola}
\affiliation{Department of Chemistry, Saint Louis University, St. Louis, USA
}

\author{N.~Marwan}%
\affiliation{Potsdam Institute for Climate Impact Research, Member of the Leibniz Association, Telegraphenberg A 31
14473 Potsdam, EU
}
\affiliation{Institute of Geosciences, University of Potsdam, Karl-Liebknecht-Str.~24--25, 14476 Potsdam, EU
}

\date{\today}

\begin{abstract}
We present a fully automated method for the optimal state space reconstruction from univariate
and multivariate time series. The proposed methodology generalizes the time delay embedding
procedure by unifying two promising ideas in a symbiotic fashion. Using non-uniform delays allows the successful 
reconstruction of systems inheriting different time scales. In contrast
to the established methods, the minimization of an appropriate cost function determines the embedding
dimension without using a threshold parameter. Moreover, the method is capable of detecting 
stochastic time series and, thus, can handle noise contaminated input without adjusting parameters.
The superiority of the proposed method is shown on some paradigmatic models and experimental data
from chaotic chemical oscillators.

\end{abstract}

\maketitle

\section{\label{sec_introduction}Introduction}

\begin{figure}
  \includegraphics[width=\linewidth]{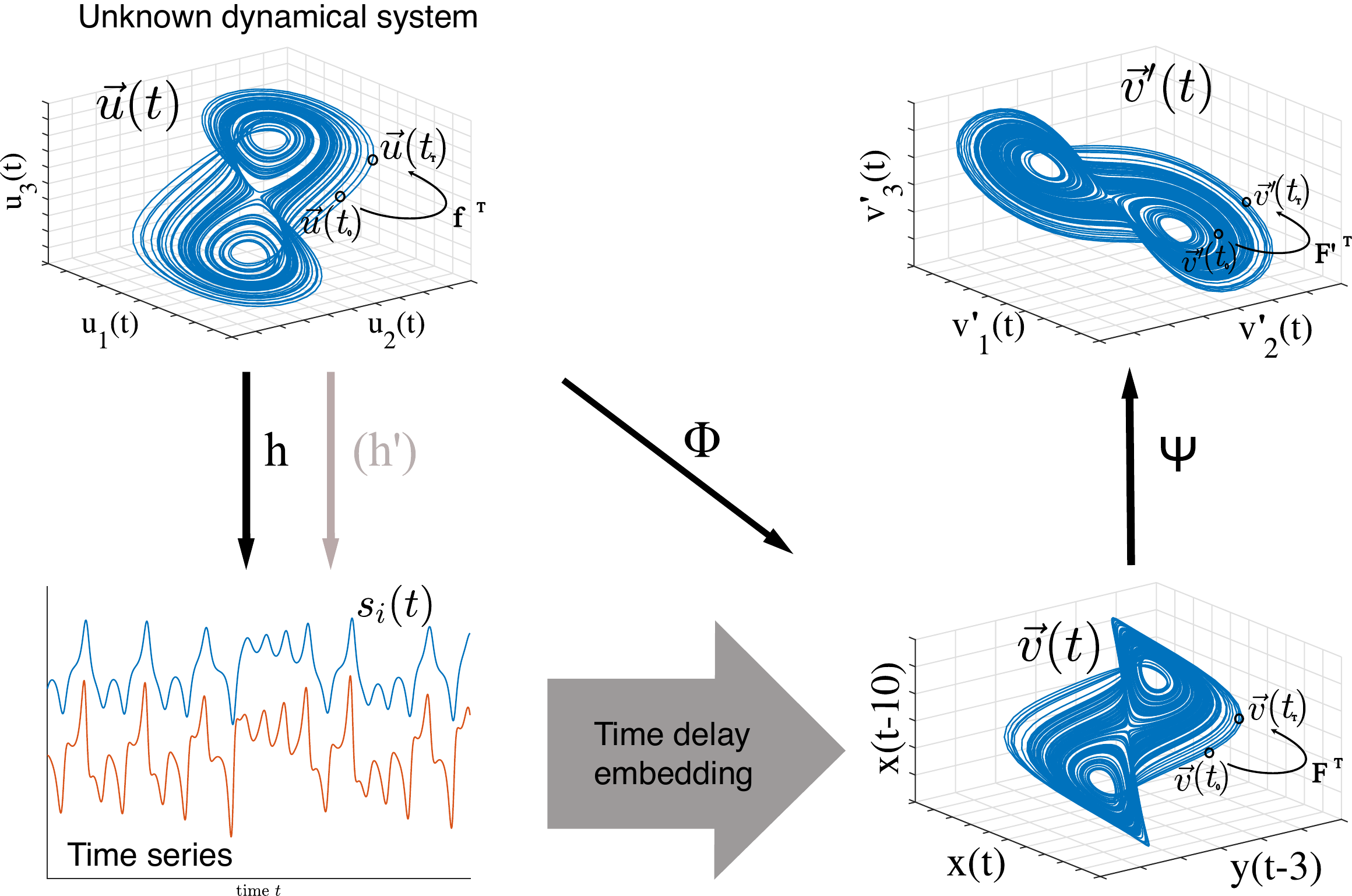}
  \caption{Schematic representation of the embedding/reconstruction procedure. See the text for details. This figure
  is inspired by \citet{casdagli1991} and \citet{uzal2011}.}
  \label{fig_embedding_scheme}
\end{figure}

State space reconstruction from observational time series often marks the first and basic step
in nonlinear time series analysis. Several methods addressed the reconstruction problem, but none of them
allow for a fully automatized and reliable way of embedding a uni- or multivariate set of observed time series with no, 
or at least very few, free parameters. The aim of this paper is to provide such a technique. 
The embedding theorems of \citet{whitney1936}, \citet{mane1981}, and \citet{takens1981} among with their extension by 
\citet{sauer1991} allow several approaches to tackle the
reconstruction problem. Among using derivative coordinates \cite{broomhead1986,mann2011}, PCA \cite{packard1980} or Legendre 
coordinates \cite{gibson1992}, uniform- and nonuniform time delay embedding \cite{packard1980} is by far the most
commonly used technique, due to its appealing simplicity. However, since Takens' theorem \cite{takens1981} is based on
noise-free and infinitely long data, it does not give any guidance to choose the proper time delay(s) $\tau$ in practice.
Together with the unknown box-counting dimension $D_B$ of the observed, but unknown system, which is needed to fulfil the 
embedding dimension criterion $m \geqslant 2D_B+1$,
the majority of the published articles propose ideas to infer estimates for $\tau$ and the reconstruction dimension
$m$ from data, usually a univariate time series (univariate embedding). 
The reconstruction problem starts with the unknown system $\vec{u}(t)$ with a mapping $f:\mathbb{R}^{D_B}\to\mathbb{R}^{D_B}$, 
which is observed via a measurement function $h$ and
lead to $M$ observational time series $\{s_i(t)|i=1,\ldots,M\}$ (Fig.~\ref{fig_embedding_scheme}).
There can be different measurement functions $h'$ forming
the multivariate dataset $s_i(t)$ and the combination of Whitney's and Takens' embedding 
theorems allow for constructing $\vec{u}(t)$ from more than one time series
(multivariate embedding) \cite{sauer1991,deyle2011}.
One then tries to find optimal embedding parameters $m$ and
$\tau$'s (the delays can be integer multiple of some constant, \textit{uniform time delay embedding \textit{UTDE}}, or different for each embedding dimension, 
\textit{non-uniform time delay embedding NUTDE}) in order to build reconstruction vectors $\vec{v}(t)$ and, thus, a mapping 
$F:\mathbb{R}^{m}\to\mathbb{R}^{m}$.
These can be furthermore transformed by $\Psi$ into $\vec{v}'(t)$ and $F'$, preserving the diffeomorphism to $\vec{u}(t)$. For a detailed
introduction into the reconstruction problem we refer to \citet{casdagli1991},
\citet{gibson1992}, \citet{uzal2011} or \citet{nichkawde2013}. 
 
In time delay embedding the key idea is to use lagged values of the available time series as components of the reconstruction
vector
\begin{equation}
\label{eq_time_delay_embedding}
\vec{v}(t) = \bigl( s_{i_1}(t-\tau_1), s_{i_2}(t-\tau_2),\ldots,s_{i_m}(t-\tau_m) \bigl).
\end{equation}
The delays $\tau_j$ and the corresponding time series $s_{i_j}, j=[1,\ldots,m]$ should be chosen, such 
that the resulting components of the
vectors forming the reconstructed state space $\vec{v}(t)$ are as independent as possible \citep{packard1980, sauer1991}, but 
at the same time preserve the correlation structure of the system to a certain extent. These two competing
objectives are also known as the problems of \textit{redundancy} (delays should not be too small) and \textit{irrelevance} 
(delays should not be too large) \cite{casdagli1991, rosenstein1994, eftekhari2018, uzal2011} and its optimization is goal 
to any proposed time delay embedding procedure.   

Despite of its lack of a sound theoretical foundation for higher dimensional reconstructions ($m>2$)
\cite{grassberger1991,vlachos2010,fraser1989}, in a univariate scenario (i.e., $s_{i_1}=\ldots=s_{i_m}=s(t)$ in Eq.~\eqref{eq_time_delay_embedding}), 
the approach to choose $\tau_2$ from the first minimum of the auto-mutual information \cite{fraser1986,liebert1989} is most common. 
$\tau_1$ is usually set to zero, i.e. the unlagged time series constitutes the first component of the reconstruction vectors.
The embedding dimension $m$ is then separately determined, usually by a false nearest neighbour 
approach \cite{kennel1992,kennel2002,cao1997,hegger1999,krakovska2015} or some other 
neighbourhood-preserving-idea \cite{aleksic1991,cenys1988}
and all delays up to $\tau_m$ are simply integer multiples of $\tau_2$ (UTDE). 
Other approaches
for an appropriate choice of $\tau_2$ are 
possible \cite{kember1993,cao1998,kugiumtzis1996,albano1988,albano1991,rosenstein1994,aguirre1995,broomhead1986}.
We refer to this as standard \textit{time delay embedding (TDE)} in the following.
More sophisticated 
ideas \cite{liebert1991,buzug1990,buzug1992a,buzug1992b,kember1993,gao1993,gao1994,judd1998,kim1999,vlachos2010,small2004,gao1994,perinelli2018}, 
some including non-uniform delays and the extension to multivariate input 
data \cite{holstein2009,nichkawde2013,garcia2005a,garcia2005b,pecora2007,hirata2006,han2019} have been presented
\cite{hirata2019,han2019}, but it seems their use is rather limited and not very popular. This could be due to their occasionally
very complex nature and the lack of high quality open source implementation in the most commonly used programming languages.
Another reason could be the fact that standard TDE performs surprisingly well in a range of examples; but still, 
its limitations should not be neglected, in particular when it comes to noisy time series, systems exhibiting multi-time scale behaviour, or multivariate
input data. The latter are becoming of increasing interest in the near future, since acquisition costs for sensors and data collection
decrease rapidly. Moreover, the application of complex systems approaches and nonlinear dynamics in different
scientific disciplines receives increasing popularity.

Here we propose a fully automated method for an appropriate state space reconstruction of uni- or multivariate time series
input, which utilizes two concepts, the \textit{continuity statistic} \cite{pecora2007} and the 
\textit{$L$-statistic} \cite{uzal2011}.
We briefly review the basic ideas we will use in our proposed method (Sect.~\ref{sec_review}) in order to illustrate
their specific utilization in the algorithm (Sect.~\ref{sec_method}), before applying it to simulated and
experimental data (Sect.~\ref{sec_applications}).

\section{\label{sec_review}Review of used concepts}

In order to ensure comprehensibility of our proposed method in Section \ref{sec_method} we explain the
two main concepts of it in the following. In Section \ref{sec_pecora_method} we review the 
\textit{continuity statistic} \cite{pecora2007} rather detailed, while the \textit{$L$-statistic} is
described only briefly in Section \ref{sec_uzal_method} and extensively in the Appendix \ref{appendix_uzal_method}.

\subsection{\label{sec_pecora_method}Continuity statistic by Pecora et al.}

In the \textit{continuity statistic}, the problem of finding an optimal state space reconstruction with 
respect to \textit{redundancy} and \textit{irrelevance} is addressed by 
a statistical determination of functional independence among the components of the reconstruction vector \cite{pecora2007}. Let 
$\{s_i(t)|i=1,\ldots,M\}$ be a multivariate dataset consisting of $M$ time series, equally sampled at time instants
$t=1,\ldots,N$. Suppose we have already chosen some delays $\tau_k$ to build our temporal reconstruction vector 
$\vec{v}(t)$ of dimension \textit{d}. This is, $\vec{v}(t)=(s_{j_1}(t+\tau_1),s_{j_2}(t+\tau_2),\ldots,s_{j_d}(t+\tau_d))$,
with $j_k\in\{ 1,\ldots,M\}$ being the choices of the different time series and $\tau_k$ the according delays, which can be
 -- and most often are -- different. Then for a new potential component of $\vec{v}(t)$ we ask if this new potential 
 component can be expressed as a function of the existing components. Mathematically speaking, the equality
\begin{equation}\label{eq_functional_independence}
s_{j_{d+1}}(t+\tau_{d+1})\stackrel{?}{=}f\left( s_{j_1}(t+\tau_1),s_{j_2}(t+\tau_2),\ldots,s_{j_d}(t+\tau_d)\right)
\end{equation}
needs to be tested in an appropriate way, i.e., a sensible choice for $f:\mathbb{R}^d \to \mathbb{R}^1$ has to be made. 
This choice can be based on the property of continuity \cite{pecora1995}.

\begin{figure}
  \includegraphics[width=.95\linewidth]{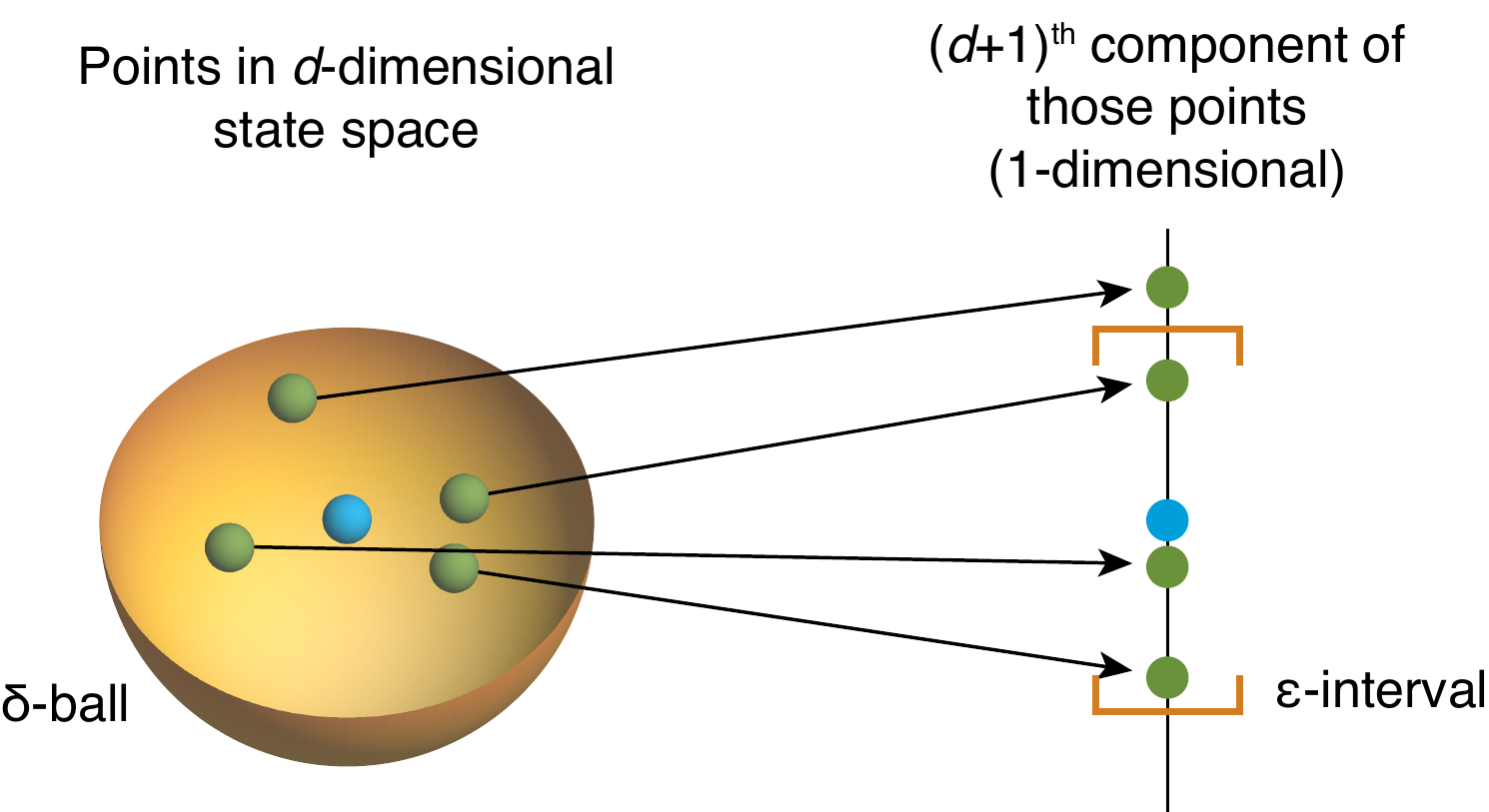}
  \caption{Fiducial point (blue) and its $k=4$ nearest neighbours (green) in the \textit{d}-dimensional $\delta$-			ball (left panel). Arrows indicate the mapping 
   $f:\mathbb{R}^d \to \mathbb{R}^1$ (right panel), 
   which is the potential $(d+1)^\text{th}$ component of each of the points in the left panel (according
   to a specific delay $\tau_{d+1}$ and time series $s_{j_{d+1}}$), Eq.~\eqref{eq_functional_independence}. 
   To decide whether this $\varepsilon$-interval size accepts or rejects the null hypothesis on a significance 			level $\alpha$
   the cumulative binomial distribution for getting at least $l=3$ points in the 
   $\varepsilon$-interval with probability \textit{p} is used (modified after \citet{pecora2007}).  
   }
  \label{fig_continuity}
\end{figure}
The practical implementation of Eq.~\eqref{eq_functional_independence} would start with mapping $k$ nearest neighbours,
$\vec{v}_{k}(t)$, of
a fiducial point $\vec{v}_\text{fid}(t)$ from {$\mathbb{R}^d \to \mathbb{R}^1$}, as illustrated in Fig.~\ref{fig_continuity}.
That is, for each of the ($k+1$) \textit{d}-dimensional points in the left panel a potential (\textit{$d+1$})th component 
$s_{j_{d+1}}(t+\tau_{d+1})$ is considered and drawn onto the number line (right panel). 
The \textit{continuity statistic}
now asks whether these $k+1$ points on the 1-dimensional number line fall within a certain $\varepsilon$-interval size by
chance, or due to the fact that there is a functional relationship between the $d$ and the (\textit{$d+1$})th component of a
potential reconstruction vector $\vec{v}(t)$. The according continuity null hypothesis is that $l$ of the $k+1$ points
landed in the $\varepsilon$-interval by chance, with probability $p$ on the basis of the binomial distribution. 
When the number of observed neighbours, which are mapped into the $\varepsilon$-interval, is larger than 
the expected number from the binomial distribution for a selected $\alpha$, i.e. $l$ points, 
then the null can be rejected and, thus, a functional relationship can be assumed. 
The number of 
considered nearest neighbours $k$ (i.e., the size of the $\delta$-ball in Fig.~\ref{fig_continuity}) also 
determines the acceptable number of $k+1-l$ 
points falling outside the chosen $\varepsilon$-interval for a given probability parameter $p$ of the binomial 
distribution and a given $\alpha$. For each candidate delay
$\tau_{d+1}$ and each time series $s_{j_{d+1}}$ for each $k$ (at a given $p$ and $\alpha$) there is a minimal spatial 
scale $\varepsilon^\star{'}$ for which the null hypothesis can be rejected, i.e., a minimal
size of the $\varepsilon$-interval in the right panel of Fig.~\ref{fig_continuity}. For the sake of avoiding
\textit{redundancy} while choosing the right delay, an $\varepsilon^\star{'}$ for each
potential $\tau_{d+1}$ has to be found. This is simply the distance from the fiducial point to its 
$l$th-nearest neighbour.
By averaging over a range of fiducial points we
eventually get the final continuity statistic $\langle \varepsilon^\star \rangle(\tau)$ as a function of 
considered delay
values $\tau$ (Fig.~\ref{fig_algorithm}).

The final idea for achieving an optimal embedding is a sequential one. For each embedding cycle $\mathcal{D}_d$, 
i.e. for trying to find an appropriate additional component to build a reconstruction vector $\vec{v}(t)$ of dimension $d+1$, initially
starting with a 1-dimensional time series $\vec{v}(t) = \{s_i(t)|i=1,\ldots,M\}$, the $\langle \varepsilon^\star \rangle(\tau)$ 
values for a range of possible delay 
values $\tau$ and for each time series $s_i(t)$ gets computed. 
The $\tau$ of the highest relative maximum of $\langle \varepsilon^\star \rangle(\tau)$ 
is selected as the optimal delay for this embedding dimension \textit{d}.
This delay is used to build up the temporal reconstruction vectors $\vec{v}(t)$ with the according time series.
From here the next embedding
cycle $\mathcal{D}_{d+1}$ gets executed and the process gets repeated until some break 
criterion terminates the procedure, i.e., when a sufficiently high embedding dimension $m$ is reached.

Even though the idea of the continuity statistic is indeed promising, in this approach several
unanswered questions remain, making the proposed idea not suitable for a fully automated embedding approach.
 
\textit{i)} The choice of $p=0.5$ has been made plausible and also our tests support this idea, while 
$\alpha=0.05$ or $\alpha=0.01$ is standard in science (see 
Figs.~\ref{fig_appendix_dependence_others}, \ref{fig_appendix_dependence_alpha}, \ref{fig_appendix_dependence_p} in 
Appendix~\ref{appendix_dependency_on_parameters}), 
so we can safely fix them to these values. What is not so clear, but highly relevant for the method, 
is how to choose the optimal
delay $\tau$ from the continuity statistic. Specifically, we might ask what ``relative maximum'' exactly means and if there is
any objective criteria for that choice. Moreover, it is also not clear how to obtain the continuity statistic in the
first place with respect to the size of the neighbourhood, i.e. the size of the $\delta$-ball in Fig.~\ref{fig_continuity}. 
We propose to vary $k$ from $k=8$ to some higher value, like $k=14$, for each considered 
delay $\tau$ and take the minimum of all trials $\varepsilon^\star{'}$ (and finally average over all fiducial points 
in order to obtain $\langle \varepsilon^\star \rangle$). This is allowed, because there is no preferred choice of $k$,
but a lower bound (see Table 1 in \citet{pecora2007}), and generally the choice depends on the amount of available data and
its sampling rate.

\textit{ii)} In the original study, it was proposed that the continuity statistic on its own provides a breaking criterion for the
method, namely, when ``$\langle \varepsilon^\star \rangle$ remains small out to large $\tau$, we do not need to
add more components.'' \cite{pecora2007} However, this is no objective criterion and introduces a statistic, which would quantify \textit{small},
and also a threshold, which determines when \textit{small} is \textit{small enough}. Due to folding and
stretching of the attractor for high delay values $\tau$, especially in case of chaotic systems, we expect $\langle \varepsilon^\star \rangle$ 
to increase with higher $\tau$, anyway. For 
these cases a (computationally intensive) irrelevance measure, the \textit{undersampling statistic}, 
has been proposed \cite{pecora2007}. Nevertheless, even though the 
undersampling statistic prevents the choice of \textit{irrelevant} delays, it does not tell which
of the local maxima of the continuity statistic we should pick and when to stop adding more components to the
reconstruction vectors \cite{comment_breaking_criterion}. 
As an alternative to assess the irrelevance, the \textit{$L$-statistic} has been suggested \cite{uzal2011} which will
be later used for our automated approach to attractor reconstruction.

\subsection{\label{sec_uzal_method}$L$-statistic by Uzal et al.}

The \textit{$L$-statistic} is an objective cost function, which quantifies the goodness of 
a reconstruction, independent of the reconstruction method \cite{uzal2011}. It has two free parameters,
$k$ and $T_M$. The approach uses ideas of 
\textit{noise amplification} and minimization of the complexity of the reconstruction, which lead to a
variable $\sigma$, and combines it with an irrelevance measure $\alpha$. Specifically, the method estimates 
the conditional variance of neighbourhoods consisting of $k$ nearest neighbours as the relative dispersal of 
these neighbourhood points with respect to the center of mass of that neighbourhood $T$ time steps ahead. 
Eventually this conditional variance estimate gets averaged over a range of prediction horizons $T$ up to a maximum 
value $T_M$ and is normalized with respect to the original neighbourhood size, thus defining $\sigma$. 
The irrelevance measure $\alpha$ is basically the averaged inter-point distance, which depends on the sampling. The
final statistic is then defined as 
\begin{equation}
	L_k = \log_{10}(\alpha_k\sigma_k) ,
\end{equation}
where the index $k$ indicates the dependence on the chosen number of nearest neighbours. A detailed description can 
be found in Appendix \ref{appendix_uzal_method}.
The authors showed, that the $L$-statistic converges for any $k\geq3$. Our analysis 
(Fig.~\ref{fig_appendix_dependence_others} in Appendix~\ref{appendix_dependency_on_parameters})
supports this assumption and, thus, we can fix $k=3$. However, the second free parameter $T_M$ will alter the resulting
$L$-statistic at any value. Particularly in the way we want to utilize the concept of this cost function in our
automated embedding scheme, we need to tackle this parameter dependency (see Sect. \ref{sec_method}). 
It is worth mentioning that the $L$-statistic inherits the minimization of a 
mean squared prediction error (MSE) (here computed using a local constant model based on the first $k$ neighbours)
and the FNN-statistic proposed by \citet{kennel1992} (when $k=1$ and $T_M=\tau$).

\section{\label{sec_method}New reconstruction method}

The $L$-statistic (Sect.~\ref{sec_uzal_method}, Appendix~\ref{appendix_uzal_method}) on its own could guide the reconstruction 
problem on finding the optimal delay values and a sufficiently high embedding dimension, when used in a brute-force-approach, i.e., 
scanning all possible delay values of all available time series in every single possible combination. It is not clear a priori
how to set the parameter $T_M$ for obtaining the $L$-statistic, so the described procedure has to be repeated for a 
range of values for $T_M$. In most cases, 
this is not computationally feasible and, therefore, not suitable for a fully automated embedding approach. We propose to combine
the \textit{continuity statistic} (Sect.~\ref{sec_pecora_method}) and the
\textit{$L$-statistic} (Sect.~\ref{sec_uzal_method}). The continuity statistic $\langle \varepsilon^\star \rangle$ 
guides the preselection of potential delays $\tau$ and time series $\{s_i(t)|i=1,\ldots,M\}$ in each embedding cycle $\mathcal{D}_d$, 
whereas the $L$-statistic decides which one to pick.

\begin{figure}[htbp]
  \includegraphics[width=\linewidth]{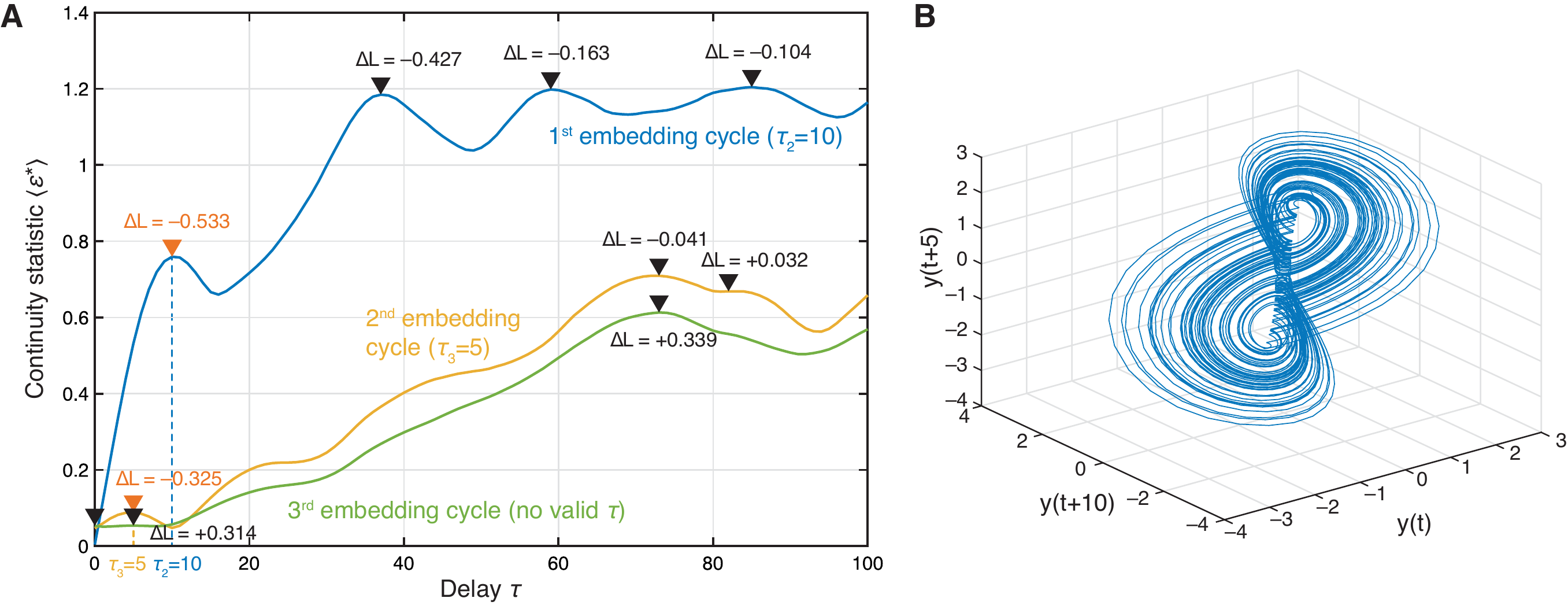}
  \caption{Illustration of the proposed embedding procedure for a univariate case, using the $y$-component of the Lorenz
  system (Appendix \ref{appendix_lorenz}). (A) Blue, yellow, and green lines represent the continuity statistics $\langle \varepsilon^\star \rangle$ for the 
  three embedding cycles the Algorithm \ref{algorithm} computes. Triangles identify the $\tau$ values corresponding to local maxima of 
  $\langle \varepsilon^\star \rangle$. Then, the local maximum which corresponds to the 
  maximum decrease of the $L$-statistic with respect to the actual reconstruction vectors
  $\vec{v}_\text{act}$, denoted as $\Delta L$ (orange triangle) is chosen as 
  a delay time in each embedding cycle.  
  In the third embedding cycle the cost-function cannot be minimized any further, i.e. all peaks lead to a
  non-negative $\Delta L$. In this case no delay value is chosen 
  and the algorithm breaks. (B) We end up
  with a 3-dimensional embedding and lags $\tau_1=0, \tau_2=10, \tau_3=5$.}
  \label{fig_algorithm}
\end{figure}

This algorithm works recursively. An embedding cycle $\mathcal{D}_d$ determines the optimal time delay and its corresponding time series 
and enables us to build the actual reconstruction vectors $\vec{v}_\text{act}$ from this, having dimension $d+1$. 
Algorithm \ref{algorithm} and Fig.~\ref{fig_algorithm}
explains the method in detail, which we refer to as ``PECUZAL'' algorithm in honour of its roots.

\begin{figure}[htbp]
  \includegraphics[width=\linewidth]{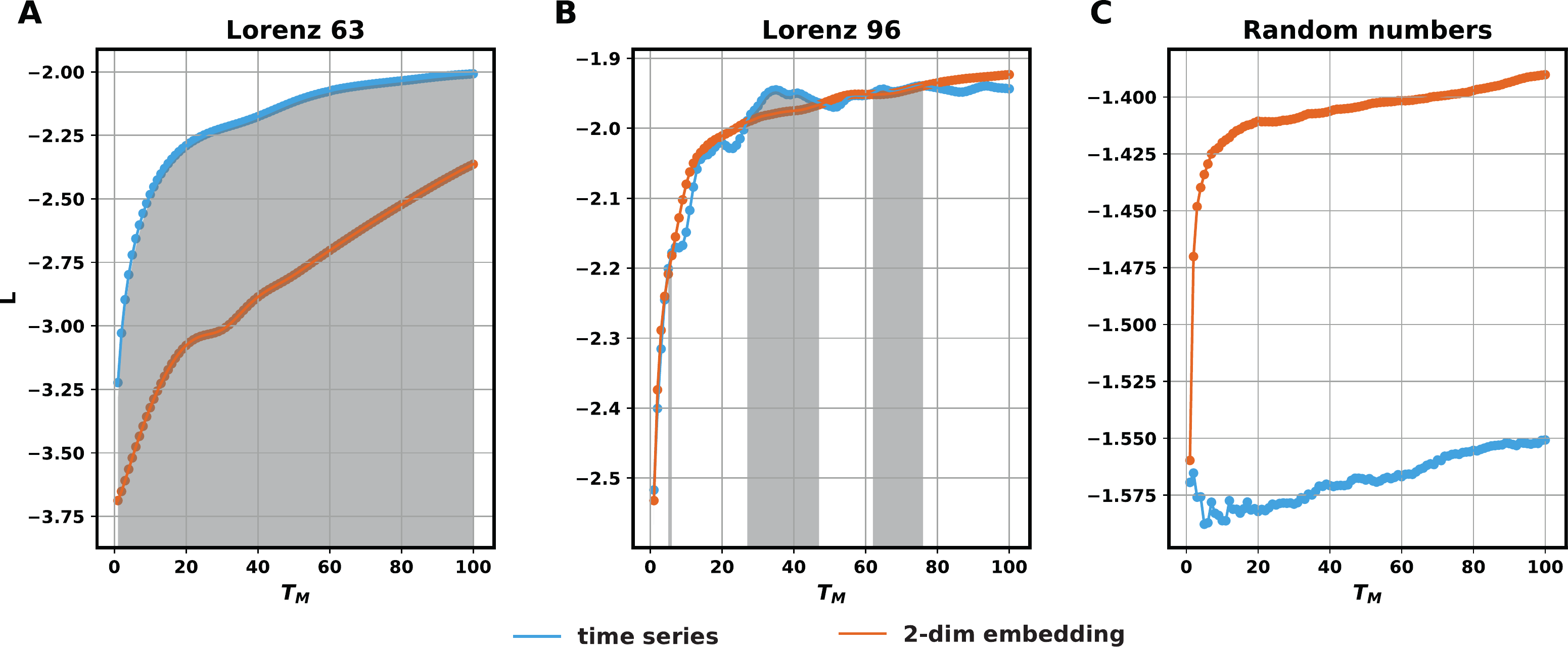}
  \caption{Illustration of the determination of $\Delta L$ within the embedding procedure for the first embedding cycle in a univariate case, 
  using (A) the $y$-component of the Lorenz system (Appendix \ref{appendix_lorenz}), (B) the time series of the second node of a $N=8$ 
  Lorenz 96 setup (Appendix \ref{appendix_lorenz96}) and (C) 1,000 uniformly distributed random numbers. Shown are the $L$-statistics for the 
  single time series (blue graphs) and a 2-dimensional embedding with a $\tau$-lagged version of itself (orange graphs) for a range of parameter 
  values $T_M$. We set the number of nearest neighbours, which constitute a neighbourhood, necessary for computing the $L$-statistic, to 
  $KNN = 3$ and estimate $\tau$ from the first minimum of the corresponding auto-mutual information. The possible decrease of the $L$-statistic 
  for this hypothetical embedding cycle with the chosen $\tau$ is simply $\Delta L(T_M) = L_{\text{orange}}(T_M) - L_{\text{blue}}(T_M)$. When 
  $\Delta L(T_M)<0$ (shaded areas) the additional reconstruction vector component does increase the quality of the reconstruction, whereas 
  when $\Delta L(T_M)>0 ~\forall~ T_M$ a further embedding is not beneficial. As expected that is the case for the stochastic signal in panel C. The
  Algorithm \ref{algorithm} automatically picks the first minimum $\Delta L$ over all $T_M$, which has also been the global minimum with respect to all $T_M$ throughout all examples we have considered so far.}
  \label{fig_dependence_Tw}
\end{figure}

{\parindent0pt
(1) For the actual reconstruction vectors $\vec{v}_{\text{act}}$, in each embedding cycle $\mathcal{D}_d$, 
$\langle \varepsilon^\star \rangle_i(\tau)$ is computed for all available $M$ time series $s_i(t)$. 
We comment on the procedure in the first embedding cycle $\mathcal{D}_1$ further below.\\}
(2) We consider all those delays $\tau_j$ for each $\langle \varepsilon^\star \rangle_i(\tau)$, which correspond 
to local maxima $\hat{\tau_j}$ in $\langle \varepsilon^\star \rangle_i(\tau)$. 
These delays $\tau_j$
(and their corresponding time series $s_{i_j}(t)$) are used to build temporary reconstruction
vectors $\vec{v}_\text{temp}(\tau_j,s_{i_j})$, by concatenating $\vec{v}_\text{act}$ with the $\tau_j$-lagged
time series $s_{i_j}(t)$. \\
(3) For each $\vec{v}_{\text{temp}}(\tau_j,s_{i_j})$ and the actual reconstruction vectors 
$\vec{v}_{\text{act}}$ the $L$-statistic is simultaneous computed for many parameters $T_M$ 
(c.f. Fig.~\ref{fig_dependence_Tw}), and we denote them as
$L_{\vec{v}_{\text{temp}}(\tau_j,s_{i_j})}(T_M)$ and $L_{\vec{v}_{\text{act}}}(T_M)$. 
We compute the maximum $L$-decrease for
$\vec{v}_{\text{temp}}(\tau_j,s_{i_j})$ with respect to $\vec{v}_{\text{act}}$ as 
$\Delta L_{{\text{temp}}(\tau_j,s_{i_j})}= \min\limits_{T_M}[L_{\vec{v}_{\text{temp}}(\tau_j,s_{i_j})}(T_M)-L_{\vec{v}_{\text{act}}}(T_M)]$. This way $T_M$ is not a free
parameter anymore.\\
(4) The delay-time series combination $(\tau_{j'},s_{i_{j'}}(t))$, which yields the minimum $\Delta L$ value 
will be picked for this embedding cycle $\mathcal{D}_d$, if $\Delta L<0$, and is used to construct the actual reconstruction vectors 
$\vec{v}_{\text{act}} = \vec{v}_{\text{temp}}(\tau_{j'},s_{i_{j'}})$ of dimension $d+1$ \\
(5) $\vec{v}_{\text{act}}$ is passed into the next embedding cycle $\mathcal{D}_{d+1}$.\\
(6) We repeat steps (1) to (5) until we cannot minimize the $L$-statistic any further, i.e. when each considered
potential embedding $\vec{v}_{\text{temp}}$ will yield a positive $\Delta L$. In this case the reconstruction
cannot get any better and we stop. $\vec{v}_{\text{act}}$ constitutes the final reconstruction.
Thus, the $L$-statistic provides a break criterion, without the introduction of 
any threshold parameter. Each embedding cycle ensures the maximum possible decrease of the cost-function.

\makeatletter
\let\OldStatex\Statex
\renewcommand{\Statex}[1][3]{%
  \setlength\@tempdima{\algorithmicindent}%
  \OldStatex\hskip\dimexpr#1\@tempdima\relax}
\makeatother

\begin{algorithm}[H]
\caption{Pecuzal Embedding}
\begin{algorithmic}[1]
\INPUT A uni- or multivariate dataset consisting of $M$ time series $s_i$ with same length and sampling and
a range of possible delay values $\tau = 0:\tau_{max}$
\State Normalize all $M$ time series to zero mean and unit variance \label{algorithm}
\State Set $\Delta L_{\text{min}} = -\text{inf}$ 
\While{$\Delta L_{\text{min}} < 0$}
	\If{1st embedding cycle $\mathcal{D}_1$}
		\State Compute $\langle \varepsilon^\star \rangle_{ik}(\tau)$ for all $M^2$ pairwise combinations of $s_i,  s_k$ for the given $\tau$'s
		\For{each peak $\hat{\tau_j}$ in each $\langle \varepsilon^\star \rangle_{ik}(\tau)$}
			\State Create $\vec{v}_{\text{temp}}$ by appending the time series $s_i$ with the $\tau_k$-lagged time series $s_k$
			\State Compute the $L$-statistics for $\vec{v}_{\text{temp}}$ and $s_i$ for a range of parameter values
			$T_M = 2:\tau_{max}$, 
			\Statex denote them as $L_{\text{temp}}(T_M)$ and $L_{s_i}(T_M)$
			\State Compute $\Delta L_{\text{temp}} = \min\limits_{T_M}[L_{\text{temp}}(T_M)-L_{s_i}(T_M)]$
		\EndFor
			\State From all $\Delta L_{\text{temp}_j}$ take the $\tau_j$, which corresponds to the peak with minimum $\Delta L$, $\Delta L_{\text{min}}$
		\State Save $\Delta L_{\text{min}}$ and $\vec{v}_{\text{temp}}$
	\ElsIf{$\mathcal{D}_d$}
		\State Compute $\langle \varepsilon^\star \rangle_i(\tau)$ for $\vec{v}_{\text{act}}$ and all $s_i$ for the given $\tau$'s
		\For{each peak $\hat{\tau_j}$ in each $\langle \varepsilon^\star \rangle_{i}(\tau)$}
			\State Create $\vec{v}_{\text{temp}}$ by appending $\vec{v}_{\text{act}}$ with the $\tau_i$-lagged time series $s_i$
			\State Compute the $L$-statistics for $\vec{v}_{\text{temp}}$ and $\vec{v}_{\text{act}}$ for a range of parameter values
			$T_M = 2:\tau_{max}$, 
			\Statex denote them as $L_{\text{temp}}(T_M)$ and $L_{\text{act}}(T_M)$
			\State Compute $\Delta L_{\text{temp}} = \min\limits_{T_M}[L_{\text{temp}}(T_M)-L_{\text{act}}(T_M)]$
		\EndFor
		\State From all $\Delta L_{\text{temp}_j}$ take the $\tau_j$, which corresponds to the peak with minimum $\Delta L$, $\Delta L_{\text{min}}$
		\State Save $\Delta L_{\text{min}}$ and $\vec{v}_{\text{temp}}$
	\EndIf
\algstore{pecuzal}
\end{algorithmic}
\end{algorithm}

\begin{algorithm}[H]
\begin{algorithmic}[1]
\algrestore{pecuzal}
	\If{$\Delta L_{\text{min}} < 0$}
		\State Set $\vec{v}_{\text{act}} = \vec{v}_{\text{temp}}$
	\EndIf
\EndWhile
\State Set $\vec{v}_{\text{final}} = \vec{v}_{\text{act}}$
\OUTPUT The final reconstruction vectors $\vec{v}_{\text{final}}$
\end{algorithmic}
\end{algorithm}

We give some remarks on the proposed idea:

\textit{i)} In case of the very first embedding cycle the actual reconstruction vectors $\vec{v}_{\text{act}}$ are not yet defined. 
Roughly speaking, the algorithm needs to find a time series to start with and which can act
as the first component of the final reconstruction vectors. As explained in Algorithm \ref{algorithm}, each of the available time series serves as 
$\vec{v}_{\text{act}}$ in the first run and consequently $M^2$ continuity statistics get computed in the first step, i.e.,
for each combination $(i, k)$ of the given time series $s_i(t)$. Note that we always use the unlagged available time series $s_i(t)$, which will
constitute the first component of the reconstruction vectors, i.e. we set $\tau_1=0$. The continuity statistic reveals the correlation structure 
of its input, meaning that a lagged initial time series would lead to different consecutive delay values. However, the difference of the finally chosen delay values
as well as the total time window of the reconstruction \textit{Tw}$= \max(\tau_1, \tau_2, \ldots)$ would be identical, because the correlation structure does not change, at least 
in case of infinite data. Practically, any $\tau_1\neq0$ only reduces the amount of data available and searching for the optimal $\tau_1$ in the sense of a
minimal $\Delta L$ for the first embedding cycle $\mathcal{D}_1$ would increase the computation time tremendously.
 
\textit{ii)} The $L$-statistic can not serve as an absolute measure, due to its dependence on the parameter $T_M$. Consider a time series 
and a potential 2-dimensional embedding consisting of this time series and a $\tau$-lagged version of it. This corresponds to the first embedding
cycle $\mathcal{D}_1$ in Algorithm \ref{algorithm}. Figure \ref{fig_dependence_Tw} shows the $L$-statistic for a range of parameters $T_M$ for the single 
time series and for the potential 2-dimensional reconstruction for two deterministic systems (panels A, B) and in case of the time series being uniformly
distributed random numbers (panel C).

Whenever the $L$-statistic of the 2-dimensional reconstruction (orange graph) is lower than the one from the single time series (blue graph) an 
embedding would be beneficial 
(gray shaded areas in Fig. \ref{fig_dependence_Tw}). But this is not always the case throughout the course of $T_M$ (see panel B). 
The conclusion is, that it is not meaningful to judge a reconstruction on a single $L$-value, gained from a fixed $T_M$. 
A reconstruction is always related to a function $L(T_M)$ of the
parameter $T_M$ and it is sensible to look at relative $L$-discrepancies between two consecutive embedding cycles, namely $\Delta L$. It turns out
that some stochastic signals will yield a negative $\Delta L$ for $T_M=1$. In panel C of Fig.~\ref{fig_dependence_Tw} this is not the case, but 
the proximity of the two curves only for $T_M=1$ is decernible. This is comprehensible by recalling that the $L$-statistic inherits 
the mean squared prediction error (see Eq.~\eqref{eq_Ek2}), and a one-sample-prediction horizon is simply too short, to properly distinguish deterministic
from stochastic sources. As a consequence we compute the $L$-statistics in each embedding cycle for $T_M$-values starting with $T_M=2$.
Thus, for any embedding cycle each peak of the continuity statistic does not receive a certain absolute $L$-value, but rather a maximum 
possible decrease of $L$, $\Delta L$, with respect to the actual embedding (Fig.~\ref{fig_algorithm}). Then one simply picks the peak, which yields the 
largest decrease. We can not rule out the possibility that we could obtain a lower overall decrease in $L$ for all embedding cycles by taking a different  
``path'', i.e. not go for the maximum decrease of $L$ in each embedding cycle. This would correspond to achieving a local minimum of the cost function in 
the parameter and embedding cycle space. 

\textit{iii)} We propose to stop the embedding process, when $\Delta L > 0$ for all considered temporary reconstruction vectors $\vec{v}_\text{temp}$.
One could think about incorporating a threshold, a small negative number, e.g. $\Delta L > -0.05$, to avoid only tiny decreases of the cost function
encountered in an embedding cycle. Throughout all our computations this has not been necessary and, therefore, we dispense on such an additional parameter.  

\textit{iv)} The continuity statistic $\langle \varepsilon^\star \rangle$ itself contains information about the correlation 
structure of the data (cf.~Sect.~\ref{sec_pecora_method}), which makes it valuable in the
context of an automated embedding procedure as proposed here, especially for multivariate input. 
Not only that the first maximum most often coincides with the value we would obtain from the first
minimum of the mutual information, but the continuity statistic of two time series ``levels off'' at a
certain value range. The absolute value of $\langle \varepsilon^\star \rangle$ represents the correlation
structure of the data we are looking at and quantifies the independence from each other for a specific time
lag. This fact allows our method to pick only time series from a multivariate set, which belong together,
and, consequently, in combination with the corresponding decrease of the $L$-statistic, $\Delta L$, 
avoid stochastic signals (cf. Fig~\ref{fig_dependence_Tw}C, Table \ref{table_results_2}). 

\textit{v)} The time complexity of the proposed method is $\mathcal{O}(N\log{}N)$ as illustrated in Fig.~\ref{fig_appendix_performance}
(Appendix \ref{appendix_implementation}).

\section{\label{sec_applications}Application, Comparison \& Results}

We apply the proposed method to a range of different systems, artificial and experimental data, exemplifying
its advantage over established methods. Specifically, we compare our method to the standard \textit{time delay embedding}
(TDE). We estimate the uniform delay $\tau$ by means of the first minimum of the auto-mutual information
\cite{fraser1986} and estimate the appropriate embedding dimension by using Cao's method \cite{cao1997}. 
Specifically, we automatically select the appropriate embedding dimension, when the $E_1$-statistic reaches the
saturation value within a given threshold of the slope (we picked a slope of $<0.05$).
We also
look at two more sophisticated methods, which can also handle multivariate input, namely the method proposed by 
Garcia \& Almeida (G\&A) \cite{garcia2005a,garcia2005b,comment_garcia} and Nichkawde's method (MDOP) \cite{nichkawde2013}.
The latter mentioned methods do not come with a predefined way to terminate the embedding process. In order to terminate 
them we use Hegger's \cite{hegger1999} method of obtaining the optimal embedding dimension. Specifically, we set
a FNN-threshold of 0.05, i.e., the algorithm breaks when the normalized fraction of FNN's fall below this threshold
in order to allow the algorithm to give meaningful results in the presence of noise.
The threshold for the tolerable relative increase of the distance
between the nearest neighbours, when increasing the embedding dimension is set to 2, as suggested in 
\cite{hegger1999,kennel1992}. The threshold, which defines the 
factor of tolerable stretching for the $d_{E_1}$-statistic in case of G\&A's method is set to 10, as suggested
by the authors. 
We estimate the decorrelation time by using the first minimum of the auto-mutual information and use it as the
Theiler window \cite{theiler1986} in all approaches. In the multivariate input cases, we pick the maximum from all 
first minima of all auto-mutual information statistics. For distance computation, the Euclidean norm is used.

\subsection{\label{sec_evaluation_statistics}Reconstruction evaluation statistics}

In order to compare our approach with the established methods we need
to quantify the goodness of the embedding. For this, we will consider six statistics.
In addition to the overall decrease of the $L$-statistic, that is $\Delta L = \sum_{i=1}^{m-1} \Delta L_i$, where $m$ is the embedding dimension and
$\Delta L_i$ the corresponding $L$-decreases in the encountered embedding cycles,
we use two other statistics, which also reflect the
neighbourhood relations of the reconstruction compared to the reference. One is
the mutual false nearest neighbour statistic \cite{rulkov1995}. Instead of estimating the coupling strength/ 
degree of synchrony of two coupled oscillators, we use the statistic for assessing the similarity between the reference
(time series gained from numerical integrations) and the reconstruction:
\begin{equation}
\text{MFNN} = \frac{1}{N} \sum_{i=1}^{N}\quad \frac{\sum_{k=1}^K |\vec{v}_i - \vec{v}_{i_{k}^\text{ref}}|}{\sum_{k=1}^K |\vec{u}_i - \vec{u}_{i_{k}^\text{ref}}|}
\quad \frac{\sum_{k=1}^K |\vec{u}_i - \vec{u}_{i_{k}^\text{rec}}|}{\sum_{k=1}^K |\vec{v}_i - \vec{v}_{i_{k}^\text{rec}}|} ,
\label{eq_mfnn}
\end{equation}
where $\vec{u}_i$ are the vectors of the reference/original system, $\vec{v}_i,~ i=1,\ldots,N$ the reconstruction vectors,
$i_{k}^\text{ref}$ the indices of the $K$-nearest neighbours of index $i$ in the original system and $i_{k}^\text{rec},~k=1,\ldots,K$ the corresponding
indices measured in the reconstruction. We propose the comparison of $K$ nearest neighbours instead of just focusing on the
first nearest neighbour, in order to receive more robust results in the presence of noise. 
By sampling the data sufficiently high we allow the precise determination of quite large neighbourhoods of a fiducial point, so we
set $K=10$ in our computations. The results vary in their absolute values with different choices of $K$, but the order of rank for
the different test methods and their relative difference remain approximately constant.
$\text{MFNN}=1$ corresponds to an ideal Afraimovich diffeomorphism \cite{pecora1995,afraimovich1986}, higher values mark worse
reconstructions. The other strict criterion we propose quantifies the degree of neighbourhood-relation conservation: the
Joint Recurrence Rate Fraction (JRRF). It is based on the concept of a recurrence plot (RP), which is a 2-dimensional
representation of a dynamical system as a binary matrix \cite{eckmann87,marwan2007} (Appendix \ref{appendix_recurrence}).
JRRF measures the accordance of the recurrence plot of the reference system, $\mathbf{R}^{\text{ref}}$, and the
RP of the reconstruction, $\mathbf{R}^{\text{rec}}$).
\begin{align}
\text{JRRF} &= \frac{\sum_{i,j}^N JR_{i,j}}{\sum_{i,j}^N R_{i,j}^{\text{ref}}} , \quad \text{JRRF} \in [0 ~ 1]\\
\mathbf{JR} &= \mathbf{R}^{\text{ref}}\circ \mathbf{R}^{\text{rec}}.
\label{eq_jrrf}
\end{align}
We compute both, $\mathbf{R}^{\text{ref}}$ and $\mathbf{R}^{\text{rec}}$, by fixing the recurrence threshold
corresponding to a global recurrence rate of 8\%. This is also to ensure comparability of the recurrence quantifiers described below
\cite{kraemer2018}. Results a fairly similar for a wide range of choices of the recurrence rate we tried and the
particular choice (in our case 8\%) is not so important, since we apply them to all RP's we compare. It is, of course, also
possible to compare different recurrence plot quantifiers gained from $\mathbf{R}^{\text{rec}}$ to the ones derived from $\mathbf{R}^{\text{ref}}$ 
\cite{marwan2007}. We here choose the determinism (DET), the diagonal line length entropy (ENTR) and the recurrence time entropy (RTE)
(Appendix \ref{appendix_recurrence}).
The latter two are related to the Kolmogorov-Sinai-Entropy \cite{march2005,baptista2010}, but do not serve as straight forward estimators, 
when necessary corrections on the RP and its quantifiers are ignored \cite{kraemer2019}.

\subsection{\label{sec_artificial_data}Paradigmatic examples}

We investigate three paradigmatic chaotic systems, the R\"ossler system in the funnel regime (Appendix \ref{appendix_roessler}), 
a driven Duffing oscillator in regular motion (Appendix \ref{appendix_roessler}) and the Mackey-Glass delay equation 
(Appendix \ref{appendix_mackey}). For all systems we compare the mean values of the evaluation statistics 
from ensembles of 1,000 trajectories with different initial conditions. Table \ref{table_results_1} in Appendix 
\ref{appendix_numerical_results} summarizes all results, also including uncertainties and results for 10\% additive measurement noise.
In order to easily compare the evaluation statistics, we use the relative deviation from the reference (e.g., 
$\frac{|DET^\text{rec}-DET^\text{ref}|}{DET^\text{ref}}$), except for the MFNN and the $\Delta L$-statistic, where we use
the relative deviation from the best score (i.e., $\frac{|MFNN^\text{rec}-MFNN^\text{best}|}{MFNN^\text{best}}$). For a concise
visual presentation we use spider plots in the following and plot \textit{1 - rel. deviation}, i.e. the closer to unity the value gets,
the better the accordance to the reference or the best achieved value is.

\begin{figure}
\includegraphics[width=\linewidth]{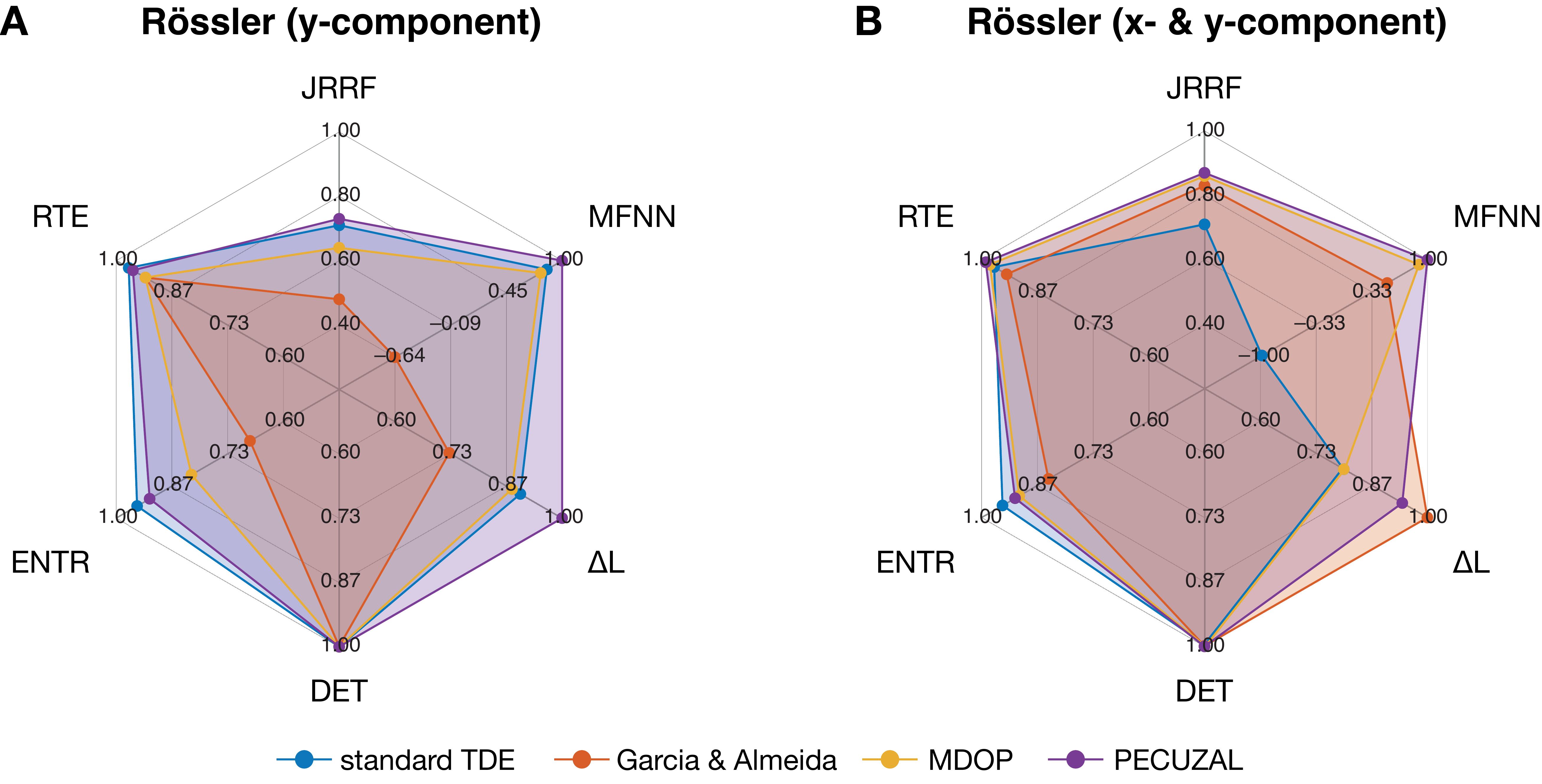}
\caption{(Relative) deviation of reconstruction by standard time delay embedding 
(TDE), the Garcia \& Almeida (G\&A), Nichkawde's (MDOP), and PECUZAL methods, comparing 
the accordance of the RPs of the reconstructed attractor and the
reference attractor (JRRF), mutual false nearest neighbors (MFNN), $\Delta L$-statistic,
as well as the recurrence quantifiers determinism (DET), diagonal line length entropy (ENTR),
and the recurrence time entropy (RTE).
(A) Univariate case using the $y$-component of the numerically integrated R\"ossler 
system (Appendix \ref{appendix_roessler}) and (B) multivariate case using the $x$- and $y$-values 
of the R\"ossler system. 
Since TDE cannot handle multivariate input we take the values from the univariate case here for illustrative
reasons, which result in different relative values in case of MFNN and the $\Delta L$-statistic. For these measures
we plot the \textit{1- relative deviations to the best score}, which increases in the multivariate case. For the
other statistics we plot \textit{1- relative deviations to the reference score}, i.e. the closer to unity the value gets,
the better the accordance to the reference or the best achieved value is.}
\label{fig_spider_roessler}
\end{figure}

\textit{(i)} 
For the chaotic R\"ossler system, we reconstruct the state space for the univariate case (using the $y$-component, in order to allow 
TDE having the best chances \cite{letellier2002}) and for the multivariate case (using the $x$- and $y$-component). An overview over the
results are shown in Figure~\ref{fig_spider_roessler}.
For TDE in the multivariate case we take the results from the univariate example, because TDE cannot handle multivariate input. Note that
this leads to different relative values for MFNN and $\Delta L$, since we plot the deviation to the best score in these cases.
The PECUZAL method performs best in the univariate as well as in the multivariate scenario, with improved outcomes for the multivariate one, as expected. 
This also holds in case of applied measurement noise, where our method even expands its lead for most measures (Table \ref{table_results_1}).
Surprisingly, TDE also provides very good results in the univariate case and in the multivariate setup Garcia \& Almeida's method yields
an overall larger deacrease of the $L$-statistic than PECUZAL, but only in the noise free case (see Table \ref{table_results_1}). This lower
$\Delta L$ is not reflected in better performance of the other evaluation statistics. Specifically, the diagonal line length entropy values 
differ from the true reference value in the double-digit percentage range for G\&A's method.

\begin{figure}[h]
\includegraphics[width=\linewidth]{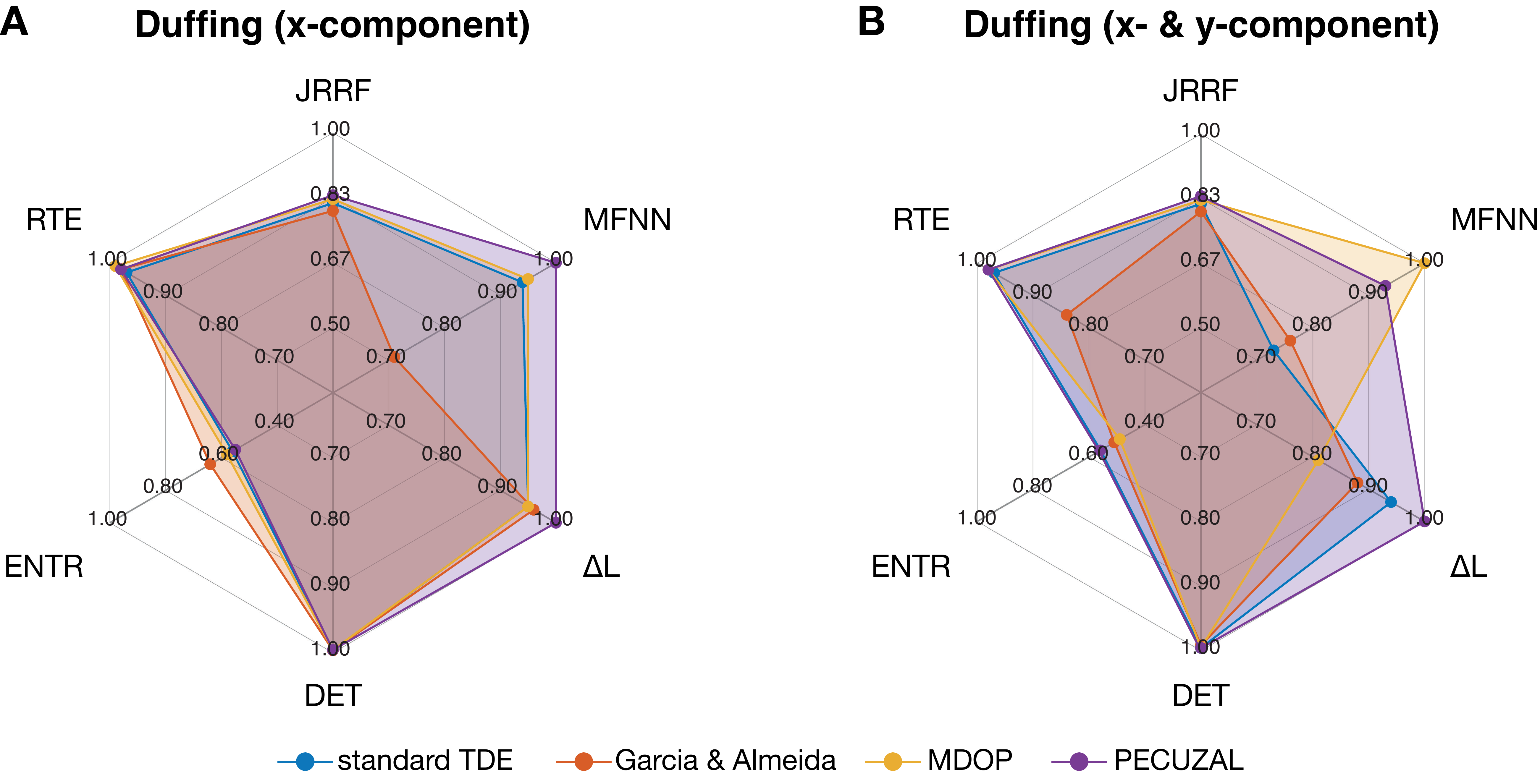}
\caption{Same as Fig.~\ref{fig_spider_roessler}, here for the Duffing system (Appendix \ref{appendix_duffing}).}
\label{fig_spider_duffing}
\end{figure}

\textit{(ii)} The overall rating also holds for the driven Duffing oscillator (Fig.~\ref{fig_spider_duffing}), but there are severe differences. 
In contrast to the chaotic R\"ossler system here it seems that the additional given time series for the multivariate scenario do not improve the
reconstructions significantly. The \textit{JRRF} do not get better and remain on a quite high level of 80\% (G\&A) up to 84\% (PECUZAL) accordance
with the reference recurrence plot $\mathbf{R}^{\text{ref}}$. This is also reflected in the $\Delta L$ statistic, which also does not improve in the
multivariate case for G\&A and MDOP, and only slightly decreases for PECUZAL. The same story is told by looking at the other evaluation statistics
summarized in Table \ref{table_results_1}. It is important to note that under noise our proposed method outperforms the others in almost all cases, but 
in principle we notice that the signal to noise ratio seems to be very low and biases the results of all methods more than in case of the R\"ossler example.
Specifically in case of \textit{RTE} deviations to the true reference value increase from the low single digit percentage range up to 62\% in the
noisy case. The reason is that in regular motion we expect a near zero value of \textit{RTE} and noise easily blurrs the diagonal lines in the
recurrence plot, leading to a broader distribution of recurrence times and, thus, to randomly elevated \textit{RTE} values. The very same problem 
make the \textit{ENTR} values deteriorate for all methods, here already apparent in the noise free case. In a regular motion system \textit{ENTR} is biased 
the most when no correction of the diagonal lines is performed, which has been shown by \citet{kraemer2019}. The proposed skeletonizaton of the recurrence
plots for the purpose of reducing the bias in the diagonal line based recurrence plot quantifiers, such as \textit{ENTR} and \textit{RTE}, lead to
way better results in the one digit percentage range even for noise, as expected. Due to the computational complexity of the skeletonization algorithm 
\cite{kraemer2019} we did not apply this correction scheme to all of the 1,000 runs in this experiment, but rather tried it on small samples not shown here, 
in order to understand the bad performances for all methods in case of \textit{ENTR} and \textit{RTE}.

\textit{(iii)} In contrast to the above systems, the Mackey-Glass system (Appendix \ref{appendix_mackey}) is infinite dimensional and we have do deal with 
a univariate time series from the numerical integration, which is why we do not have a ``reference'' value we 
could base our computations on. Therefore, we can only use the $\Delta L$-statistic (Table \ref{table_results_1}). 
The proposed PECUZAL method performs significantly better
than the other methods with Garcia \& Almeida's method coming second, especially in the presence of noise. Here all methods but the
proposed one yield reconstructions with corresponding positive $\Delta L$-values, i.e. the suggested embeddings do not decrease the overall 
$L$-statistic. Admittedly, PECUZAL also comes with a very small negative $\Delta L$ value, with 0 falling within the $1\sigma$-interval. This finding 
indicates a
too low signal to noise ratio, which we comment on below.
\citet{farmer1982} conjectured a linear relation between the
delay chosen in the Mackey-Glass equation and the corresponding dimensionality of the attractor (c.f.~Table I in \cite{farmer1982}). A
linear fit to the data of that table suggests an attractor-dimension of $d_A \approx 5.3$ with the 95\% confidence interval being 
between $\approx 2.1773$ and $\approx 8.2888$.
The studied methods give $5\pm0 ~(6.7\pm.5)$ (TDE), $3\pm0 ~(4.6\pm1)$ (G\&A), $4\pm0 ~(2\pm0)$ (MDOP) and $7\pm0 ~(2.2\pm1.7)$ (PECUZAL). The
bracketed values correspond to the case of 10\% additive noise. While all methods meet Farmer's conjecture in the noise free scenario, this does 
not hold for the MDOP method an the proposed method in the noisy case. Both methods suggest too low embedding dimensions.
This is due to the fact that the signal to noise ratio is apparently too low and PECUZAL treats the signal as a stochastic source 
for some realizations where it does not embed the data at all, while it did not do it in case of the R\"ossler and Duffing system, despite 
the same variance of the white Gaussian noise. 
Results from G\&A and TDE do fall
in the 95\% confidence interval, which is large, because of the weak data basis given in Ref.\cite{farmer1982}) and the resulting uncertain fit. 
We find the time window of the embedding, i.e., the total time span covered by a reconstruction vector, 
decreasing with increasing noise level throughout our experiments. This is very much in line with the findings of \cite{ragwitz2002}.

We finally look at two made up, ill-defined multivariate datasets, in order to see
how the G\&A, MDOP, and the PECUZAL method cope with redundant data and with stochastic signals. 

\begin{figure}[htbp]
\includegraphics[width=\linewidth]{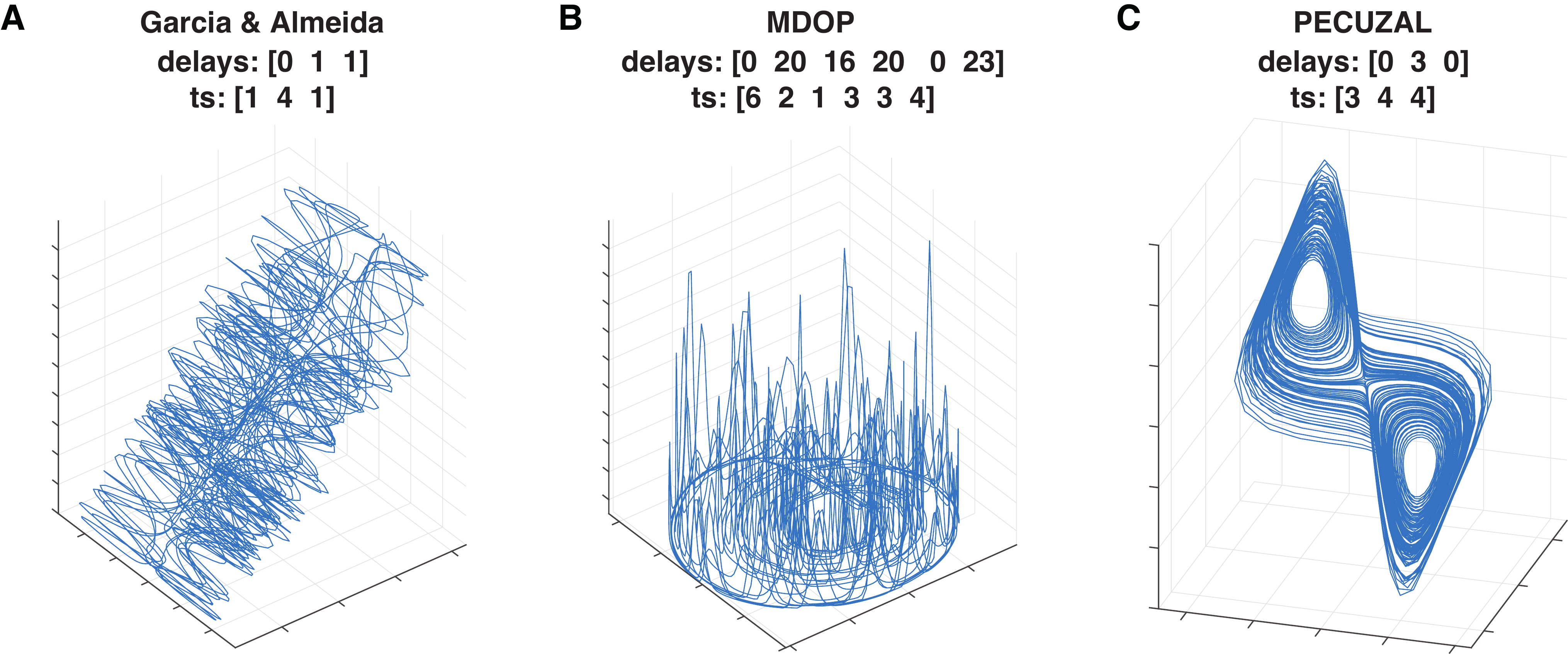}
\caption{Reconstructions of the \textit{Fooling dataset I} (see text for details). In case of the MDOP method 
(panel B), we plot the first three components of the 6-dimensional trajectory. While G\&A and
MDOP methods mix timeseries from different systems, leading to nonsensical reconstructions (panels A and B),
our proposed method (panel B) sticks to time series from one system (Lorenz in this case).}
\label{fig_fooling_1}
\end{figure}

\textit{(i)} First we construct a 
dataset consisting of six time series (\textit{Fooling dataset I}). The first two time series are the $x$- and
$y$-component of the R\"ossler system (Appendix \ref{appendix_roessler}), the third and fourth time series are the $x$- and
$y$-component of the Lorenz system (Appendix \ref{appendix_lorenz}), the fifth time series corresponds to
$x_{\text{R\"ossler}}^2 + y_{\text{R\"ossler}}$,
whereas the sixth time series is set to
$x_{\text{Lorenz}} \cdot y_{\text{Lorenz}} + y_{\text{Lorenz}}$.
Our proposed method does not mix time series from both systems and sticks to one system (Lorenz in this case) (Tab.~\ref{table_results_2}). It suggests a 3-dimensional
embedding and also does not need the redundant information stored in the fifth and the sixth time series of the
input dataset. In contrast,  G\&A and MDOP fail here, suggesting a 3-dimensional and a 
6-dimensional embedding, respectively, mixing up the different systems yielding a useless reconstruction (Fig.~\ref{fig_fooling_1}).

\textit{(ii)} The second made-up dataset (\textit{Fooling dataset II}) consists of three time series of length 5,000, with the first one being an 
auto-regressive process of order 1 with parameters (0,~0.9) and initial condition
$u_0=0.2$, to mimic coloured noise. The second and third time series are Gaussian and uniform random numbers. 
While G\&A and MDOP embed the non-deterministic time series, our proposed algorithm suggests
no embedding and throws an error (Tab.~\ref{table_results_2}). The reason is, that the $L$-statistic is a monotonically increasing function of the 
embedding dimension for stochastic data for any prediction horizon parameter $T_M$, i.e., the algorithm cannot minimize $L$ already in the first embedding cycle. 
For the sake of completeness we have to stress that this particular example should not be read as a claim of a generalizable behaviour of our proposed method to
deal with auto-regressive processes of arbitrary order $p$. In the case of higher-order AR processes, PECUZAL often suggests an embedding with a dimension that 
corresponds to the order of the AR process, as we would expect it theoretically. We have noticed, however, that the embedding depends heavily on the length of 
the time series used, which is in line with the findings of \citet{holstein2009}, but also of the choice of the particular AR-parameters and the order of the 
AR process under study. A systematic consideration of PECUZAL's embedding suggestions for this class of processes is beyond the scope of this paper.

\begin{table}[htbp]
\caption{Embedding dimension, the accordingly chosen time series and corresponding time lags for the fooling datasets, 
mimicking mixed deterministic data from different systems and redundant time series as well as stochastic time series.
We compare Garcia \& Almeida's (G\&A), Nichkawde's (MDOP), and our proposed PECUZAL method.}
\resizebox{\textwidth}{!}{%
\begin{tabular}{l@{\extracolsep{12pt}}ccc@{\extracolsep{12pt}}ccc@{\extracolsep{12pt}}ccc}

\toprule
\rule{0pt}{3ex}

& \multicolumn{3}{c}{m} & \multicolumn{3}{c}{chosen time series} & \multicolumn{3}{c}{Delay's}\\ 
\cline{2-4} \cline{5-7} \cline{8-10}

\textbf{System} & \color{black}{G\&A} & \color{black}{MDOP} & \color{black}{PECUZAL} &  \color{black}{G\&A} & \color{black}{MDOP} & \color{black}{PECUZAL} & \color{black}{G\&A} & \color{black}{MDOP} & \color{black}{PECUZAL}\\
 
\colrule 
\rule{0pt}{3ex}%

\textbf{Fooling dataset I}
& 3 & 6 & 3 & 1, 4, 1 & 6, 2, 1, 3, 3, 4 & 3, 4, 4 & 0, 1, 1 & 0, 20, 16, 20, 0, 23 & 0, 3, 0\\

\textbf{Fooling dataset II} 
& 6 & 6 & --- & 1, 2, 2, 1, 1, 1 & 3, 3, 1, 2, 2, 1 & --- & 0, 1, 2, 3, 2, 1 & 0, 1, 0, 1, 0, 1 & - \\

\botrule

\end{tabular}
}
\label{table_results_2}

\end{table}

\subsection{\label{sec_experimental_data}Experimental data}

We will now utilize the PECUZAL embedding method on experimental data. Specifically, we look at a chaotic time series from 
electrochemical oscillations. The experiment was performed with the chaotic electrodissolution of nickel 
in sulfuric acid \cite{Kiss:2000vf}. A standard three-electrode electrochemical cell was used with a 1-mm diameter nickel wire as working, a Pt counter, and a 
$\textrm{Hg/Hg}_2\textrm{SO}_4/\textrm{sat.~K}_2\textrm{SO}_4$ reference
electrode. The electrolyte was  4.5 M $\textrm{H}_2\textrm{SO}_4$ at 10\textdegree C. The nickel wire was connected to the working point of the 
potentiostat through an individual resistance ($R_{\textrm{ind}}$), and a potentiostat (Gill AC, ACM Instruments) applied a constant circuit potential 
($V_0$, with respect to the reference electrode). At a given circuit potential, the rate of the metal dissolution, measured as the current, can exhibit chaotic 
oscillations due the hidden negative differential resistance and additional nonlinear processes related to the passivation kinetics \cite{Wickramasinghe:2010ct}. 
About 500 current oscillations were recorded at a data acquisition rate of 200Hz, which corresponds to about 200 points per cycle and a time series length of $N=100,000$. 
There are two primary bifurcation parameters in the experiment: the individual resistance, which affects the charging time constant of the electrical double layer, and 
the circuit potential, which drives the dissolution. We consider a setting with $R_{\textrm{ind}}$ = 1.5~k$\Omega$  and 
$V_0$ = 1,360~mV. 

\begin{figure}[htbp]
\includegraphics[width=.8\linewidth]{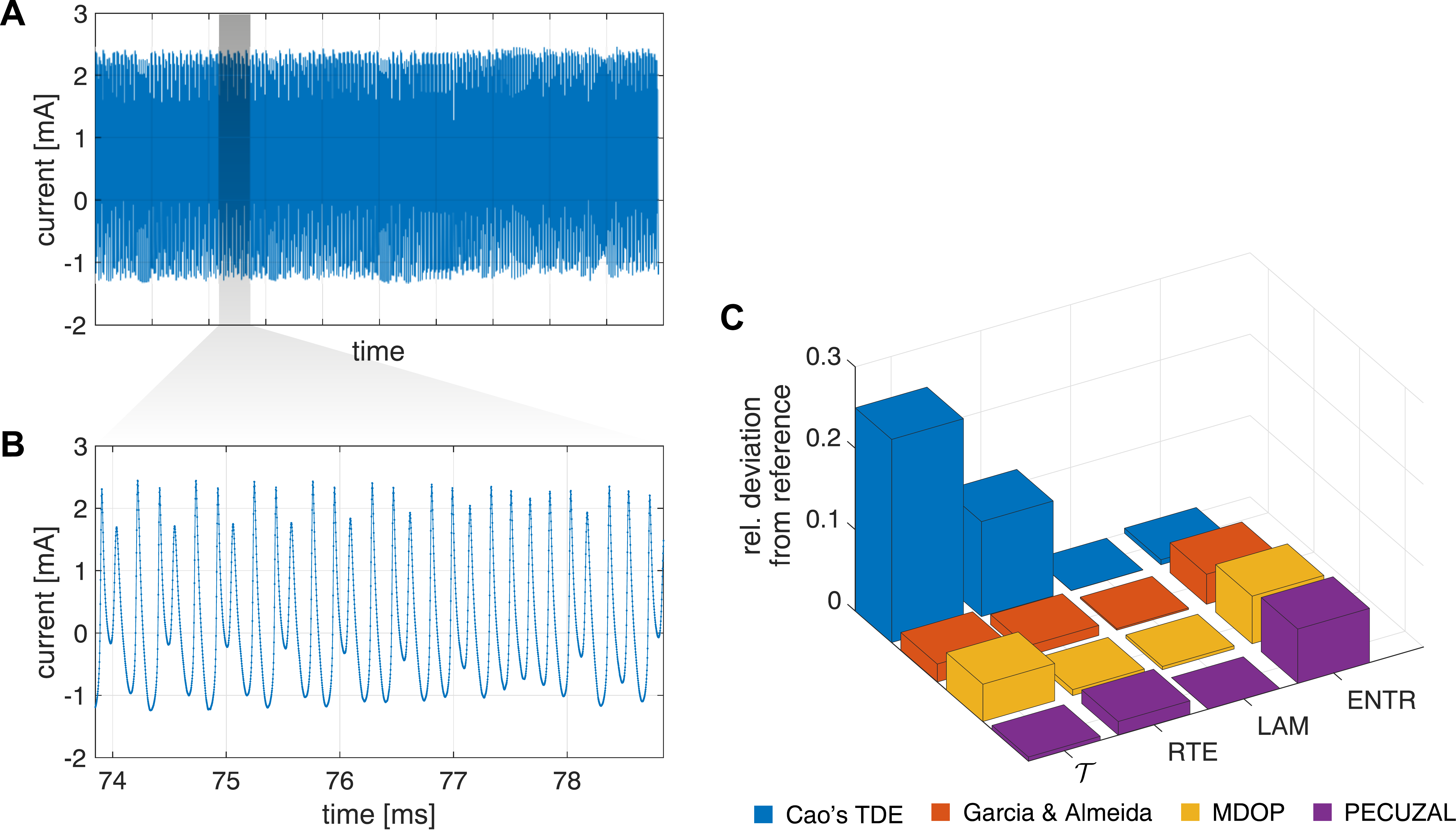}
\caption{(A) Entire z-standardized and downsampled time series from electrochemical oscillations of length $N=50,000$ and (B) a sub-sample of length $\hat{N}=2,500$. 
(C) The relative
deviation of the medians of the distributions of the RQA-quantifiers ENTR, LAM, RTE and $\mathcal{T}$ obtained from RPs of $1,000$ sub-samples to their reference 
values obtained from RPs of the entire time series for the four reconstruction methods.}
\label{fig_experimental_results}
\end{figure}

We demonstrate the ability of the proposed embedding method to cope with rather small to intermediate sized experimental datasets. We first downsample the time series
to $N=50,000$ (Fig.~\ref{fig_experimental_results}A). Reconstructions for the four methods (TDE, G\&A, MDOP, PECUZAL) are then computed from which we obtain RPs 
(with a fixed recurrence rate of 8\%, in order to guarantee comparability) and the corresponding RQA-quantifiers diagonal line length
entropy (ENTR), the laminarity (LAM), the recurrence time entropy (RTE) and the recurrence network measure transitivity ($\mathcal{T}$), see 
Appendix \ref{appendix_recurrence}. We denote each of these
values for each of the reconstruction method as its reference values. We then repeat the described procedure for $1,000$ sub-samples of length $\hat{N}=2,500$ drawn from 
the time series at random (shown exemplary in Fig.~\ref{fig_experimental_results}B), i.e. for each of the $1,000$ sub-samples we compute the reconstruction for each of 
the four methods, its corresponding RP and the RQA-quantifiers. This will result in distributions for ENTR, LAM, RTE and $\mathcal{T}$ for 
each reconstruction method. Finally we compare the medians of these distributions to their reference values and plot the relative deviation 
$\frac{|\text{RQA}_{\text{median(distr.)}}-\text{RQA}_{ref}|}{\text{RQA}_{ref}}$ in Fig.~\ref{fig_experimental_results}C. The capability of the four methods to allow
for satisfying estimates from short time series samples differs strongly for the different RQA-quantifiers. The largest discrepancies to the reference can be 
noted for TDE in case of $\mathcal{T}$ and RTE. For LAM all methods estimate the reference very well from the subsamples. While our proposed method slightly comes 
last in case of ENTR, it yields the best results for $\mathcal{T}$ and LAM and is performing well in case of RTE. The example shown here can not be generalized, 
but it underpins our claim that PECUZAL provides robust state space reconstructions for a very broad range of processes under different conditions, which are 
often better, but always at least equally well than the ones obtained from the established methods.

\section{\label{sec_conclusions}Conclusions}

A fully automated approach for state space reconstruction from uni- or even multivariate time series is 
proposed and compared to established methods.
The algorithm works iteratively and appends the reconstruction vectors in each embedding cycle with an
appropriate time delay and an according time series until a cost function cannot be minimized further.  
Its core functionality is based on identifying potential time delays and its corresponding time series in each
embedding cycle by using the continuity statistic. For each of those delays, temporary reconstruction vectors
are build and the cost function is computed. The delay value, which yields the maximum decrease of the cost function 
is selected. If none of the considered delay values yields a decrease of the cost function the reconstruction can
not get any better and the final embedding is obtained without the need of setting any threshold parameter.
Usually the time delays chosen that way are not simply multiples of each other, but rather reflect
even complex correlation structures within the data. This is why the algorithm is also able to detect time
series stemming from a stochastic source, which it will not embed. Except from computing the decorrelation
time of the data for providing a valid Theiler window for the nearest neighbour search, and providing a range of
possible delay values the algorithm shall encounter, there are neither any data
 preprocessing steps necessary, nor any free parameters need to be
adjusted before using the proposed routine. 
The approach is demonstrated on a variety of exemplary systems as well
as on experimental data stemming from chaotic chemical oscillators. 
We find that it provides often better, but always at least equally 
well reconstructions than the established methods. It is furthermore capable of providing meaningful reconstructions
for rather short time series, which particularly holds for the case of multivariate input.
The additional computational effort in
comparison to standard time delay embedding is manageable and justified. Since the proposed method works
automatically, is basically parameter free, and can handle multivariate input without mixing data originating 
from different systems, we can think of a wide range of potential applications. This is especially true for
scenarios, where multiple sensors or channels of a detector monitor real world processes, which are not
isolated observables, i.e., in engineering, earth- and life science contexts. The provided software (Appendix 
\ref{appendix_implementation}) in three common coding languages will facilitate the use of the presented method.

\begin{acknowledgments}
K.H.K thanks Dr.~Paul Schultz and Maximilian Gelbrecht for fruitful discussions and important hints and Dr.~Jaqueline Lekscha
for the idea to understand and implement the continuity statistic in the first place.  I.Z.K. acknowledges support from the 
National Science Foundation (NSF) (Grant No. CHE-1900011). This work has been financially 
supported by the German Research Foundation (DFG projects MA4759/8 and MA4759/9).
\end{acknowledgments}

\clearpage
\appendix

\section{Implementation and code availability}\label{appendix_implementation}

The study that we present here is available as a fully reproducible code base
\cite{kraemer_julia}.
In addition, we have implemented performant versions of the embedding algorithms,
as well as the automated pipeline for optimal embedding.
A single package is provided for Python \cite{kraemer_python} (https://pypi.org/project/pecuzal-embedding/) and a toolbox for 
MATLAB\textsuperscript{\textregistered} \cite{kraemer_matlab}, 
\newline (https://de.mathworks.com/matlabcentral/fileexchange/86004-pecuzal-embedding-algorithm-for-matlab) while for the Julia language
we have integrated the PECUZAL algorithm into the library DynamicalSystems.jl~\cite{datseris2018}
(the other methods ``TDE, Garcia\&Almedia, MDOP'' were already part of the library).
The automated pipeline for optimal embedding has been further refined for better user experience
and is also part of DynamicalSystems.jl.

\begin{figure}[bp]
\centering
  \includegraphics[width=.7\linewidth]{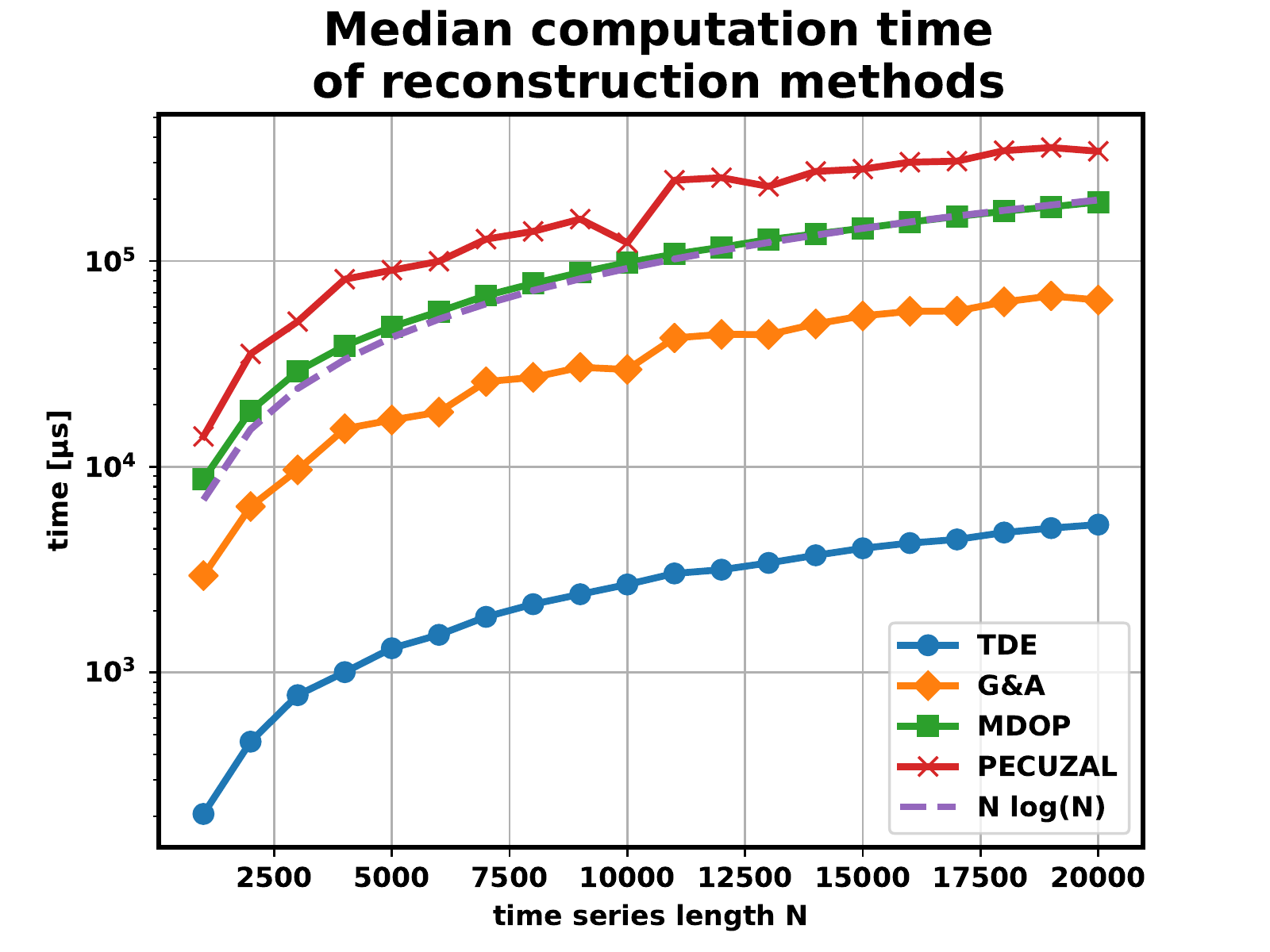}
  \caption{Median time complexity for the state space reconstruction from the $y$-component of the
  R\"ossler system (Appendix \ref{appendix_roessler}) for TDE, Garcia \& Almeida's, MDOP, and
  PECUZAL method. Ensembles of function calls for each considered time series
  length $N$ are computed and the median values are shown (implemented in BenchmarkTools.jl \cite{chen2016}).
  A Theiler window is set as the first minimum of the auto-mutual information and the maximum encountered time 
  delay is set to four times the Theiler window for TDE, G\&A, and PECUZAL.}
  \label{fig_appendix_performance}
\end{figure}

The Julia versions of the algorithms have been heavily optimized and are 
the most performant ones by orders of magnitude in parts, when compared to the 
Python (and MATLAB\textsuperscript{\textregistered}) implementation. Due to the nearest neighbour tree search,
all compared methods have $\mathcal{O}(N\log{}N)$ time complexity (Fig.~\ref{fig_appendix_performance}). 
Our proposed method
is slightly (but consistently) slower than the established methods, with standard time delay embedding
performing best (Cao's method). Worth mentioning is the fact that in the MDOP case the most computational effort 
is put on the estimation of the maximum considered delay value and not on the computation of
their $\beta$-statistic. Considering that an 
optimal embedding is a one-time operation, we believe that all methods shown here are practically
useful with respect to their computational complexity.

\section{Numerical Results}\label{appendix_numerical_results}

Chaotic R\"ossler system, driven Duffing oscillator in regular motion, and nonlinear 
time-delay Mackey-Glass equation are used to evaluate the performance of the different
embedding approaches (TDE, G\&A, MDOP, and PECUZAL).
We consider univariate and multivariate embedding (R\"ossler and Duffing). 
The $y$-component of the R\"ossler system and the $x$-component of the Duffing system are
used for the univariate embedding, wheres the $x$- and $y$-values of the corresponding systems
are used for the multivariate embedding.
For the TDE only univariate embedding is possible and, thus, there are no results
for the multivariate case. 

Ensembles of 1,000 trajectories using different initial conditions are used.
Additionally, we consider additive noise with an amplitude of 10\% of the standard 
deviation of the corresponding signal.

All RPs are computed using a fixed recurrence rate of 8\% and a minimum line length of 2.

\begin{sidewaystable}

\caption{Results of the evaluation measures: accordance of the recurrence plot (JRRF), 
mutual false nearest neighbors (MFNN), $\Delta L$-statistic, as well as the recurrence quantifiers determinism
(DET), diagonal line length entropy (ENTR), and recurrence time entropy (RTE) for the 
standard time delay embedding 
(TDE), Garcia \& Almeida (G\&A), Nichkawde's (MDOP), and PECUZAL-method 
(see text for details), using 
chaotic R\"ossler system, driven Duffing oscillator in regular motion (both in univariate (u)
and multivariate (m) case), and nonlinear time-delay Mackey-Glass equation. 
Mean values and standard deviation of ensembles consisting of 1,000 integrated trajectories are
shown. Bracketed values present results for additive noise. The results of recurrence quantifiers are
presented as relative deviations from the reference.
The best results for each considered system
are highlighted in black.}
\label{table_results_1}

\resizebox{\textwidth}{!}{%

\begin{tabular}{l>{\color{gray}}r>{\color{gray}}r>{\color{gray}}r>{\color{gray}}r@{\extracolsep{12pt}}>{\color{gray}}r>{\color{gray}}r>{\color{gray}}r>{\color{gray}}r@{\extracolsep{12pt}}>{\color{gray}}r>{\color{gray}}r>{\color{gray}}r>{\color{gray}}r}

\toprule
\rule{0pt}{3ex}

& \multicolumn{4}{c}{JRRF} & \multicolumn{4}{c}{MFNN} & \multicolumn{4}{c}{$\Delta L$}\\ 
\cline{2-5} \cline{6-9} \cline{10-13}

\textbf{System} & \color{black}{TDE} & \color{black}{G\&A} & \color{black}{MDOP} & \color{black}{PECUZAL} & \color{black}{TDE} & \color{black}{G\&A} & \color{black}{MDOP} & \color{black}{PECUZAL} & \color{black}{TDE} & \color{black}{G\&A} & \color{black}{MDOP} & \color{black}{PECUZAL}\\
 
\colrule 
\rule{0pt}{3ex}%

\textbf{R\"ossler (u)} 
& $.71\pm .02$ & $.48\pm .04$ & $.64\pm .01$ & \color{black}{$.73\pm .01$}
& $3.8\pm .3$ & $8.7\pm 2.4$ & $4.0\pm .4$ & \color{black}{$3.3\pm .2$}
& $-1.27\pm .09$ & $-1.03\pm .17$ &  $-1.24\pm .08$ & \color{black}{$-1.41\pm .08$}\\

& ($.71\pm .02$) & ($.6\pm .03$) & ($.62\pm .01$) & \color{black}{($.73\pm .01$)}
& ($4.1\pm .3$) & ($5.1\pm .6$) & ($4.8\pm .4$) & \color{black}{($3.7\pm .2$)}
& ($-1.14\pm .05$) & ($-0.94\pm .08$) & ($-0.95\pm .04$) & \color{black}{($-1.23\pm .05$)}\\

\textbf{R\"ossler (m)} 
& --- & $.83\pm .02$ & $.86\pm .04$ & \color{black}{$.87\pm .01$}
& --- & $1.9\pm .6$ & $1.4\pm .1$ & \color{black}{$1.29\pm .08$}
& --- & \color{black}{$-1.58\pm .2$} & $-1.26\pm .12$ & $-1.48\pm .08$\\

& --- & ($.82\pm .04$) & \color{black}{($.87\pm .01$)} & \color{black}{($.87\pm .01$)}
& --- & ($2.4\pm .6$) & ($1.57\pm .06$) & \color{black}{($1.5\pm .1$)}
& --- & ($-1.14\pm .09$) & ($-1.04\pm .07$) & \color{black}{($-1.28\pm .04$)}\\

\textbf{Mackey-G} 
& --- & --- & --- & ---
& --- & --- & --- & ---
& $-.43\pm .0$ & $-.57\pm .0$ & $-.35\pm .0$ & \color{black}{$-.77\pm .0$}\\

& --- & --- & --- & ---
& --- & --- & --- & ---
& ($.12\pm .04$) & ($.12\pm .07$) & ($1\times10^7\pm .0$) & \color{black}{($-0.02\pm .03$)}\\

\textbf{Duffing (u)} 
& $.82\pm .01$ & $.80\pm .04$ & $.83\pm .02$ & \color{black}{$.84\pm .02$}
& $4.0\pm 1.2$ & $4.9\pm 1.7$ & $3.9\pm .9$ & \color{black}{$3.8\pm .8$}
& $-1.66\pm 1.7$ & $-1.68\pm .18$ & $-1.66\pm .16$ & \color{black}{$-1.75\pm .15$}\\

& \color{black}{($.74\pm .01$)} & ($.63\pm .04$) & ($.69\pm .02$) & ($.73\pm .02$)
& ($8.9\pm .4$) & ($10.2\pm 1.6$) & \color{black}{($8.8\pm .3$)} & ($15.1\pm 3.7$)
& ($-.5\pm .05$) & ($-.51\pm .08$) & ($-.41\pm .06$) & \color{black}{($-.69\pm .04$)}\\

\textbf{Duffing (m)} 
& --- & $.80\pm .06$ & $.83\pm .02$ & \color{black}{$.84\pm .02$}
& --- & $3.9\pm 1.5$ & \color{black}{$3.1\pm .8$} & $3.3\pm .9$
& --- & $-1.6\pm .2$ & $-1.4\pm .2$ & \color{black}{$-1.8\pm .2$}\\

& --- & ($.72\pm .04$) & \color{black}{($.74\pm .02$)} & ($.73\pm .01$)
& --- & ($10.0\pm 1.8$) & \color{black}{($7.9\pm .6$)} & ($14.1\pm 3.9$)
& --- & ($-.5\pm .1$) & ($-.3\pm .1$) & \color{black}{($-.71\pm .04$)}\\

\end{tabular}

}

\resizebox{\textwidth}{!}{%

\begin{tabular}{l>{\color{gray}}r>{\color{gray}}r>{\color{gray}}r>{\color{gray}}r@{\extracolsep{12pt}}>{\color{gray}}r>{\color{gray}}r>{\color{gray}}r>{\color{gray}}r@{\extracolsep{12pt}}>{\color{gray}}r>{\color{gray}}r>{\color{gray}}r>{\color{gray}}r}

\toprule
\rule{0pt}{3ex}

& \multicolumn{4}{c}{DET ($\times10^3$)} & \multicolumn{4}{c}{ENTR} & \multicolumn{4}{c}{RTE}\\ 
\cline{2-5} \cline{6-9} \cline{10-13}

\textbf{System} & \color{black}{TDE} & \color{black}{G\&A} & \color{black}{MDOP} & \color{black}{PECUZAL} & \color{black}{TDE} & \color{black}{G\&A} & \color{black}{MDOP} & \color{black}{PECUZAL} & \color{black}{TDE} & \color{black}{G\&A} & \color{black}{MDOP} & \color{black}{PECUZAL}\\
 
\colrule 
\rule{0pt}{3ex}%

\textbf{R\"ossler (u)} 
& $.03\pm .01$ & $.71\pm .28$ & $.03\pm .01$ & \color{black}{$.01\pm .01$} 
& \color{black}{$.05\pm .01$} & $.32\pm .03$ & $.18\pm .01$ & $.08\pm .01$ 
& \color{black}{$.03\pm .01$} & $.07\pm .02$ & $.07\pm .01$ & $.04\pm .01$\\

& ($6.6\pm .8$) & ($7.2\pm 1.2$) & ($8.0\pm .7$) & \color{black}{($3.0\pm .4$)} 
& ($.14\pm .02$) & \color{black}{($.05\pm .03$)} & ($.08\pm .02$) & ($.08\pm.01$)
& \color{black}{($.02\pm.01$)} & ($.04\pm .01$) & ($.04\pm .01$) & \color{black}{($.02\pm.01$)}\\

\textbf{R\"ossler (m)}
& --- & $.95\pm .35$ & $.19\pm .06$ & \color{black}{$.14\pm .03$} 
& --- & $.2\pm .04$ & $.09\pm .02$ & \color{black}{$.08\pm .02$} 
& --- & $.06\pm .02$ & $.02\pm .01$ & \color{black}{$.01\pm .01$}\\

& --- & ($6.6\pm 2.4$) & ($4.9\pm .4$) & \color{black}{($2.8\pm .3$)} 
& --- & \color{black}{($.05\pm .03$)} & ($.08\pm .01$) & \color{black}{($.06\pm .01$)}
& --- & ($.02\pm .01$) & \color{black}{($.01\pm .01$)} & \color{black}{($.01\pm .0$)}\\ 

\textbf{Duffing (u)} 
& $3.9\pm .08$ & \color{black}{$3.6\pm 1.0$} & $4.1\pm .3$ & $4.2\pm .3$
& $.44\pm .09$ & \color{black}{$.36\pm .07$} & $.42\pm .02$ & $.45\pm .05$ 
& $.03\pm .04$ & $.02\pm .03$ & \color{black}{$.01\pm .01$} & $.02\pm .02$\\

& ($97.4\pm 4.3$) & ($113.8\pm 14.2$) & ($113.4\pm 15.4$) & \color{black}{($61.5\pm 5.7$)} 
& \color{black}{($.03\pm .02$)} & ($.30\pm .06$) & ($.37\pm .05$) & ($.21\pm .09$)
& ($.62\pm .01$) & \color{black}{($.32\pm .16$)} & ($.54\pm .03$) & ($.46\pm .06$)\\ 

\textbf{Duffing (m)}
& --- & \color{red}{$8.3\pm 19$} & \color{black}{$4.2\pm .3$} & \color{black}{$4.2\pm .3$}
& --- & $.49\pm .12$ & $.51\pm .07$ & \color{black}{$.44\pm .06$} 
& --- & \color{red}{$.16\pm .57$} & \color{black}{$.02\pm .02$} & \color{black}{$.02\pm .03$}\\

& --- & ($113\pm 18$) & ($103\pm 10$) & \color{black}{($61\pm 4$)}
& --- & ($.32\pm .07$) & \color{black}{($.22\pm .07$)} & \color{black}{($.22\pm .08$)}
& --- & ($.56\pm .09$) & ($.61\pm .01$) & \color{black}{($.44\pm .06$)}\\ 

\botrule
\end{tabular}

}

\end{sidewaystable}

\clearpage

\section{Dependency on parameters}\label{appendix_dependency_on_parameters}

\begin{figure}
\centering
  \includegraphics[width=\linewidth]{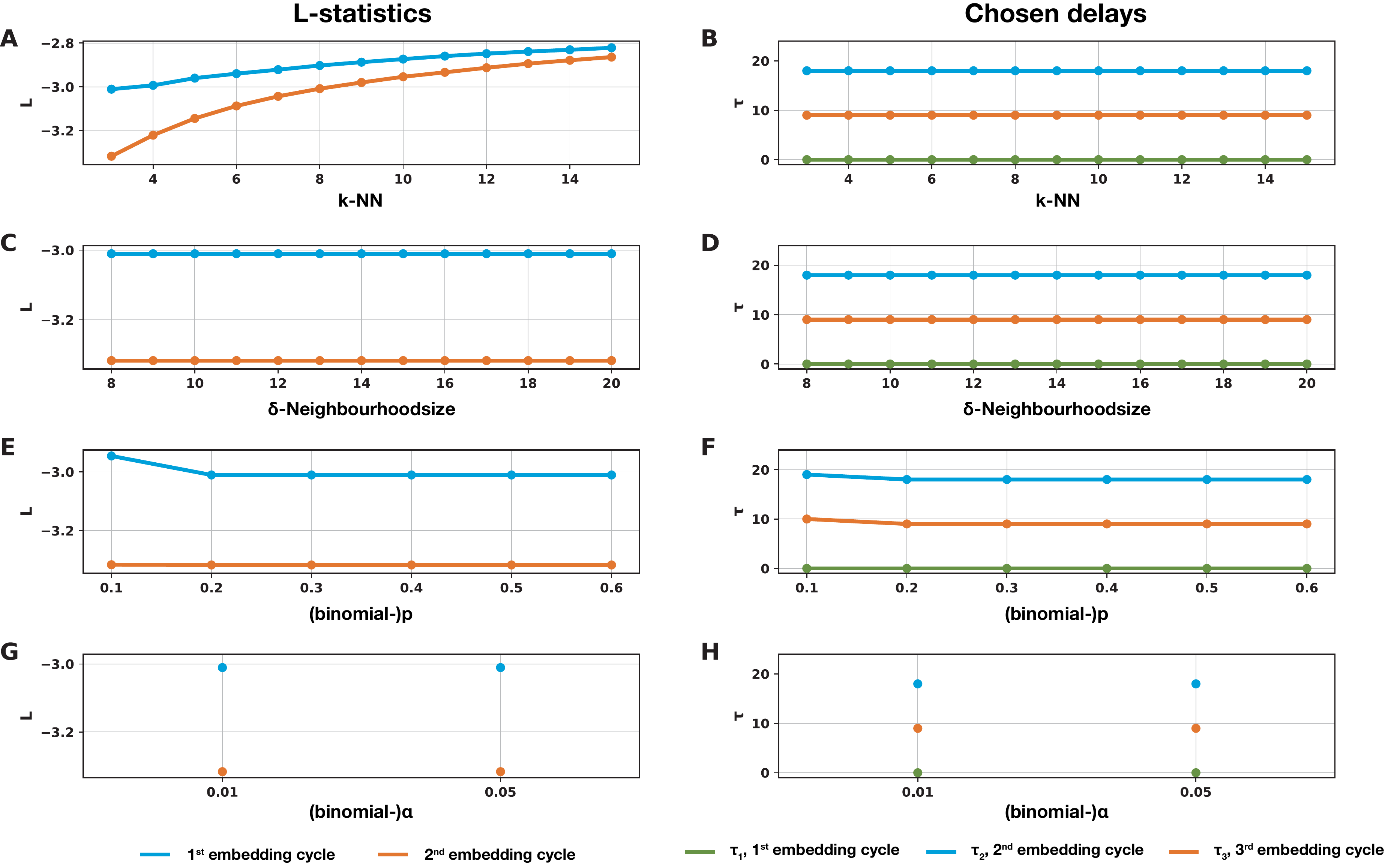}
  \caption{$L$-statistic (left panels) and chosen delays (right panels) for a variety of choices of
  parameters for embedding the Lorenz system, relevant for the PECUZAL method. See text for details.
  When not varied, the other parameters were fixed to $\delta$-Neighbourhoodsize$ =14$, $p=0.5$, 
  $\alpha=0.05$ and $kNN=3$, as we propose.}
  \label{fig_appendix_dependence_others}
\end{figure}

\begin{figure}
\centering
  \includegraphics[width=\linewidth]{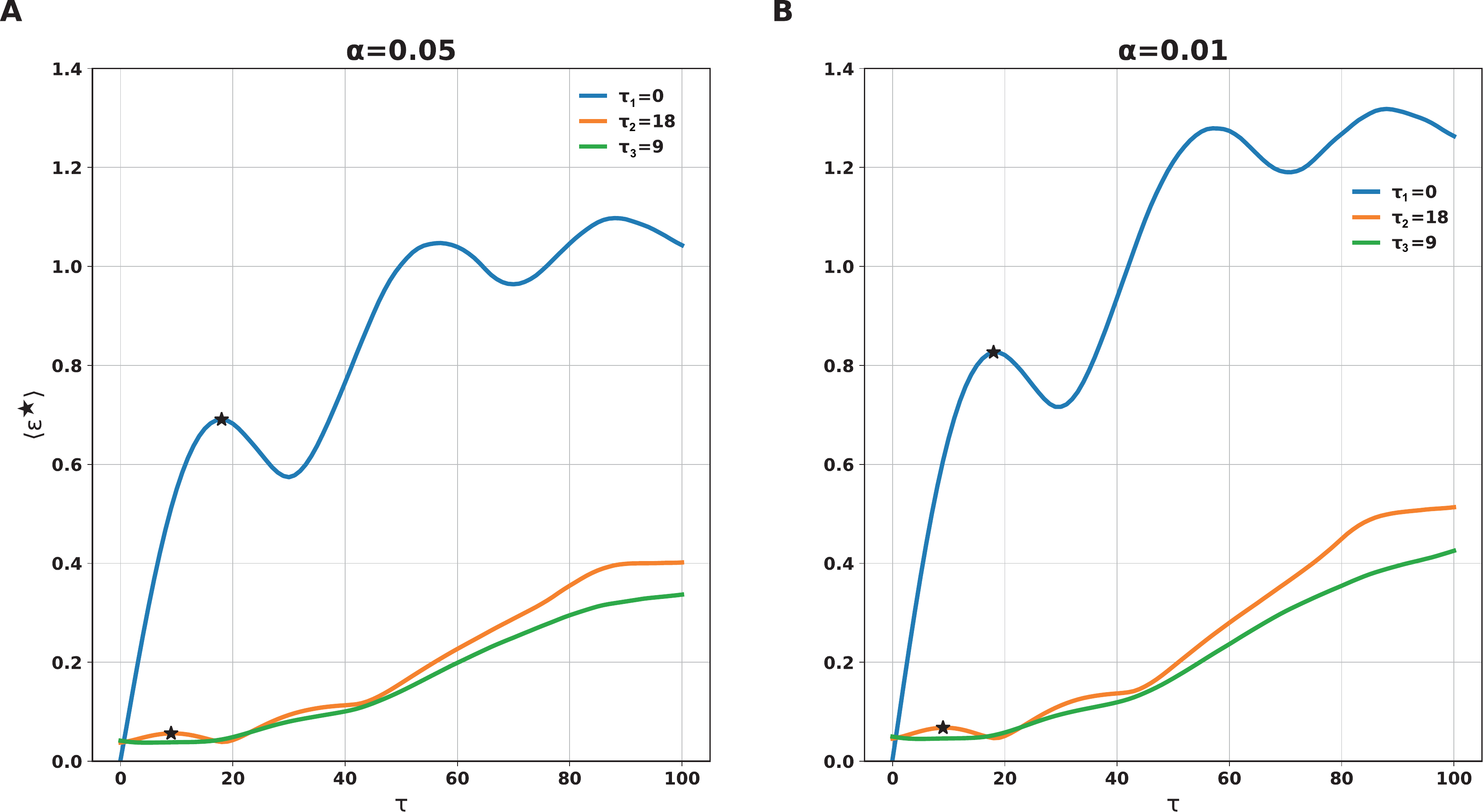}
  \caption{Impact of different significance-level choices $\alpha$ on the continuity statistic. The other parameters for
  obtaining the continuity statistic were fixed to $\delta$-Neighbourhoodsize$ =14$ and $p=0.5$.}
  \label{fig_appendix_dependence_alpha}
\end{figure}

\begin{figure}
\centering
  \includegraphics[width=\linewidth]{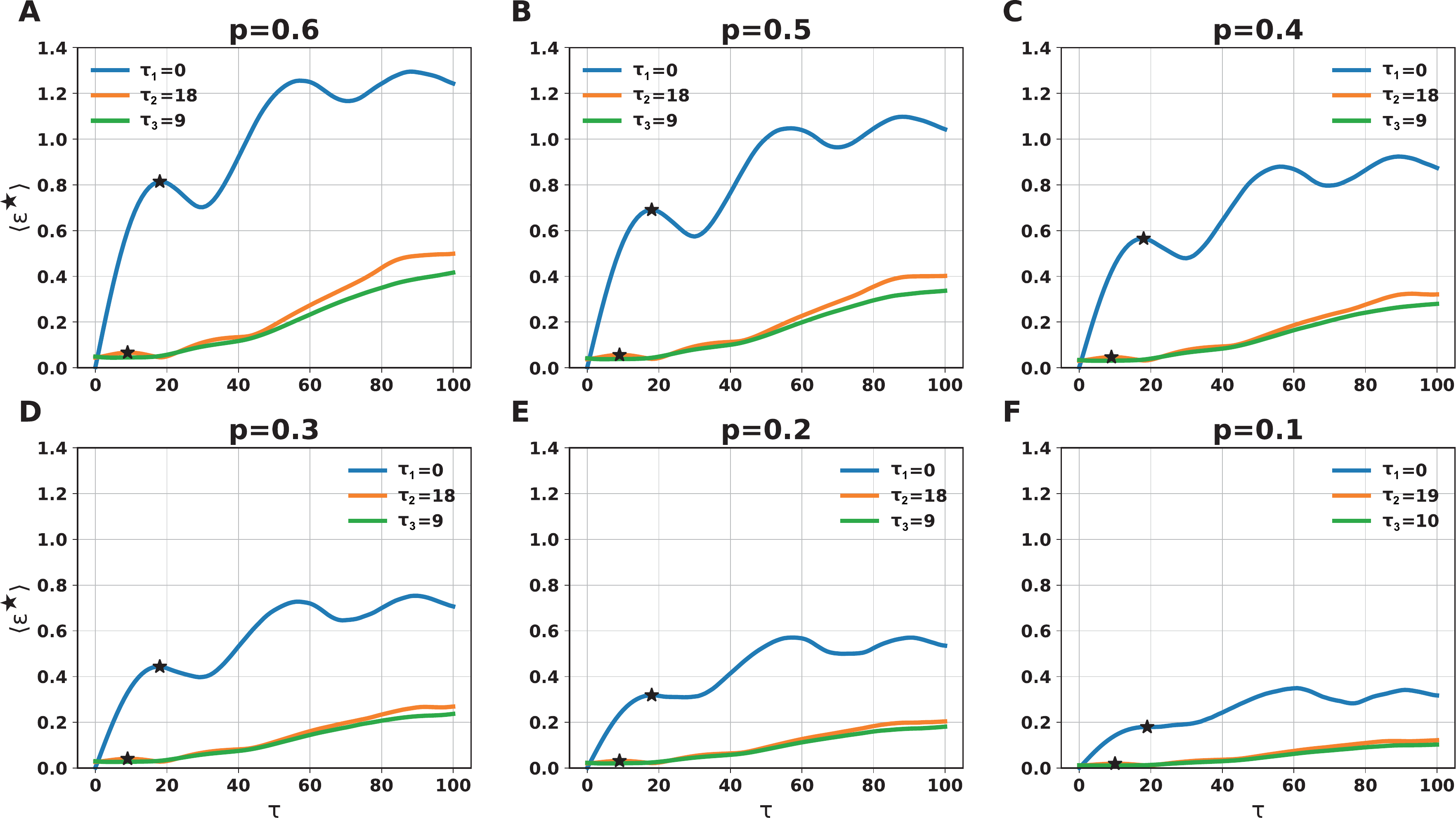}
  \caption{Impact of different choices of the binomial probability parameter $p$ on the continuity statistic. The other parameters for
  obtaining the continuity statistic were fixed to $\delta$-Neighbourhoodsize$ =14$ and $\alpha=0.05$.}
  \label{fig_appendix_dependence_p}
\end{figure}

We investigate the impact of different parameter settings on the resulting
reconstruction of our proposed method for the $x$-component of the Lorenz system 
(Appendix \ref{appendix_lorenz}). 
In particular, Figure~\ref{fig_appendix_dependence_others}) shows the sensitivity of the $L$-statistic value 
and the chosen delays with respect to the number of nearest neighbours $kNN$ for
a fixed paramter $T_M = 20$ (panels A, B) and also the dependence on
the continuity statistic parameters $\delta$-Neighborhoodsize (panels C, D), 
the (binomial) probability $p$ (panels E, F) and significance-level $\alpha$ (panels G, H).
The critical dependence of the $L$-statistic on the parameter $T_M$ is discussed in Sect.~\ref{sec_method}
and Fig.~\ref{fig_dependence_Tw}. The results show very little impact on the reconstruction quality and, 
thus, confirm our choice of fixed parameter values for the algorithm. 
Specifically, the crucial qualitative course of the continuity statistic, i.e., 
the position of the local maxima, remains unchanged for relevant choices of $\alpha$ and in the vicinity of $p=0.5$,
which is the proposed fixed value (Figs.~\ref{fig_appendix_dependence_alpha} and \ref{fig_appendix_dependence_p}).

\section{\label{appendix_recurrence}The recurrence plot and its quantification measures}

A recurrence plot (RP) is a binary, square matrix $\mathbf{R}$ representing the recurrences of states $\vec{x}_i$ ($i=1,\ldots,N$, with $N$ 
the number of measurement points) in 
the $d$-dimensional state space \cite{eckmann87,marwan2007}
\begin{equation}\label{eq_rp_definition}
{R}_{i,j}(\varepsilon) = \Theta\left(\varepsilon - \| \vec{x}_i - \vec{x}_j\|\right), \qquad \vec{x} \in \mathbb{R}^d,
\end{equation}
with $\|\cdot \|$ a norm, $\varepsilon$ a recurrence threshold, and $\Theta$ the Heaviside function. There are several ways
to quantify the structures and dynamics encoded therein, but here we look at the determinism \cite{webber1994}, which
is the fraction of recurrence points that form diagonal lines of length $\ell$
\begin{equation}\label{eq_det}
\text{DET} = \frac{\sum_{\ell=\ell_{\min}}^N P(\ell)}{ \sum_{\ell=1}^{N} \ell P(\ell)},
\end{equation}
with $P(\ell)$ being the histogram of all diagonal line lengths in the RP and $\ell_{\min}$ the considered minimal
line length (here set to 2). Another quantifier is the Shannon
entropy of the probability distribution $p(\ell) = \frac{P(\ell)}{\sum_{\ell} P(\ell)}$
to find a diagonal line of exact length $\ell$
\begin{equation}\label{eq_S_RQA}
\text{ENTR} = -\sum_{\ell=\ell_{\min}}^{N} p(\ell) \ln p(\ell).
\end{equation}

It is also possible to look at the Shannon entropy of the length distribution of the white vertical lines $\ell_\text{w}$, 
which correspond to recurrence times
\begin{equation}\label{eq_S_RTE}
\text{RTE} = -\sum_{\ell_\text{w}=1}^{N} p(\ell_\text{w}) \ln p(\ell_\text{w}).
\end{equation}

Analogue to the definition of the determinism, the laminarity is the fraction of recurrence points that form vertical lines
\begin{equation}\label{eq_lam}
\text{LAM} = \frac{\sum_{\ell_\text{v}=\ell_{\text{v},\text{min}}}^N P(\ell_\text{v}v)}{ \sum_{\ell_\text{v}=1}^{N} \ell_\text{v}v P(\ell_\text{v}v)},
\end{equation}
with $\ell_v$ the length of a vertical line, $P(\ell_\text{v}v)$ the histogram of all vertical lines in the RP and $\ell_{\text{v},\text{min}}$ the considered minimal
line length (here set to 2). 

Finally, the RP can be considered as the adjacency matrix of an $\varepsilon$-recurrence network \cite{zou2019}, $\mathbf{A} = \mathbf{R}-\mathbb{1}$ and define the transitivity 
\cite{boccaletti2006} as
\begin{equation}\label{eq_trans}
\mathcal{T} = \frac{\sum_{i,j,k=1}^N A_{jk}A_{ij}A_{ik}}{\sum_{i,j,k=1}^N A_{ij}A_{ik}(1-\delta_{jk})},
\end{equation}
which is a measure characterising the geometric structure of the state space attractor \cite{donner2011}.

\section{Exemplary models}

The models described in the following are all integrated using DynamicalSystems.jl and DifferentialEquations.jl 
\cite{datseris2018,rackauckas2017} in the Julia Language.

\subsection{Lorenz system}\label{appendix_lorenz}

The Lorenz system \cite{lorenz1963} is defined as
\begin{align*}
\dot{x} &= \sigma(y-x) \\
\dot{y} &= x(\rho-z)-y \\
\dot{z} &= xy-\beta z .
\end{align*}
We set the initial condition to $u_0=[1.0, 1.0, 50.0]$, use a sampling time of $\Delta t=0.01$ and discard the first 1,000 points of the
integration as transients. For producing Fig.~\ref{fig_algorithm} we set the parameters to 
$\sigma=10, \beta=8/3, \rho=60$ and use a time series consisting of 5,000 samples. In 
Figs.~\ref{fig_dependence_Tw}, \ref{fig_appendix_dependence_others}, \ref{fig_appendix_dependence_alpha},
 \ref{fig_appendix_dependence_p} and the redundancy fooling-dataset in section \ref{sec_artificial_data} we use the standard parameter 
 values $\sigma=10,\ \beta=8/3,\ \rho=28$ and also a time series consisting of 5,000 samples.
 
\subsection{Lorenz 96 system}\label{appendix_lorenz96}

The Lorenz 96 system \cite{lorenz1996} is defined as
\begin{align*}
\frac{dx_i}{dt} = (x_{i+1} - x_{i-2}) x_{i-1} - x_i + F
\end{align*}
with $x_i$ the state of the system for nodes $i = 1,\dots,N$ and it is assumed that the total number of nodes is $N\geq4$. The forcing constant
$F$ serves as the control parameter. Here we set $N = 8$, $F = 4.472$ and the initial condition to 
$u_0=[0.590; 0.766; 0.566; 0.460; 0.794; 0.854; 0.200; 0.298]$, use a sampling time of $\Delta t=0.1$ and discard the first 2,500 points of the
integration as transients leaving us with a time series consisting of 5,000 samples.
 
\subsection{R\"ossler system}\label{appendix_roessler}

The R\"ossler system \cite{roessler1976} is defined as
\begin{align*}
\dot{x} &= -y-z \\
\dot{y} &= x+ay \\
\dot{z} &= b+ z(x-c) .
\end{align*}
We randomly choose the initial conditions uniformly from an interval $[0,\,2]$ and discard the first 2,000 
points of the integration as transients working with time series of length $N = 10,000$ (for Fig.~\ref{fig_appendix_performance} we 
set $N = 20,000$). The parameters are set to $a = 0.2925,\ b = 0.1,\ c = 8.5$ (funnel regime) for the experiment in Sect.~\ref{sec_artificial_data}, where 
the sampling time is set to $\Delta t=0.03$ and $a = 0.2,\ b = 0.2,\ c = 5.7$ with a sampling time of $\Delta t=0.02$ elsewhere.

\subsection{Driven Duffing oscillator}\label{appendix_duffing}

The driven Duffing/Van der Pol oscillator \cite{duffing1918,appleton1922} is defined as
\begin{align*}
\dot{x} &= y \\
\dot{y} &= \mu(1-x^2)y - \alpha x - \beta x^3 + z \\
\dot{z} &= \gamma \cdot \omega \cdot \cos(\omega\cdot t),
\end{align*}
with $\mu = 0.1,\ \alpha=1,\ \beta=0,\ \gamma=0.5,$ and $\omega = 2$ \cite{cui2018} resulting in
a regular, quasi periodic motion. We randomly choose the initial conditions $x_0,\ y_0,\ z_0$ uniformly from the interval 
$[0,\,0.2]$, use a sampling time of $\Delta t=0.1$ and discard the first 2,000 
points of the integration as transients, resulting in time series of length $N = 5,000$.

\subsection{Mackey-Glass equation}\label{appendix_mackey}

The Mackey-Glass equation \cite{mackey1977} is the nonlinear time delay
differential equation
\begin{align*}
\dot{x} = \beta \frac{x_{\tau}}{1+x^n} -\gamma x,
\end{align*}
with the lag $\tau=44$, and the parameters $n =10$, $\beta=0.2$, and $\gamma = 0.1$. $x_{\tau}$ 
represents the value of $x$ at time $t$-$\tau$. 
We randomly choose the initial conditions uniformly from an interval $[0,\,1.5]$, use a sampling time of 
$\Delta t=0.5$ and discard the first 2,000 
points of the integration as transients, resulting in time series of length $N = 10,000$.

\section{Details of the $L$-statistic}\label{appendix_uzal_method}

The concept of quantifying noise amplification in the context of the validation of an attractor reconstruction has
been proposed in \cite{casdagli1991}. The reconstruction process is considered in the presence of noise and the finite 
data availability as a modeling problem, introducing a noise amplification and estimation/prediction error for any 
measures on the reconstructed attractor. The variance of the conditional probability density function is a
``natural criterion for assessing predictability'' \cite{casdagli1991} and is used as the noise amplification for a
fiducial point $\vec{v}_\text{fid}(t)$ at a given noise level $\varepsilon$ on the reconstruted attractor
\begin{equation}\label{eq_noise_amplification1}
\sigma_{\varepsilon}\bigl(T,\vec{v}_\text{fid}(t)\bigr) = \frac{1}{\varepsilon}\sqrt{Var\Bigl(\vec{v}\bigl(t+T\bigr)|B_{\varepsilon}\bigl(\vec{v}_\text{fid}(t)\bigr)\Bigr)},
\end{equation}
with $Var\Bigl(\vec{v}\bigl(t+T\bigr)|B_{\varepsilon}\bigl(\vec{v}_\text{fid}(t)\bigr)\Bigr)$ being the conditional variance of $\vec{v}(t+T)$ for
$\vec{v}_\text{fid}(t)$ in a radius $\varepsilon$ ball $B_{\varepsilon}\bigl(\vec{v}_\text{fid}(t)\bigr)$ for a prediction
horizon $T$. Finally, the noise amplification $\sigma$
\begin{equation}\label{eq_noise_amplification2}
\sigma\bigl(T,\vec{v}_\text{fid}(t)\bigr) = \lim\limits_{\varepsilon \to 0}\sigma_{\varepsilon}\bigl(T,\vec{v}_\text{fid}(t)\bigr)
\end{equation}
averaged over all fiducial points on the attractor (and squared), $\langle \sigma(T) \rangle^2$, serves as a 
measure of the predictive power, with respect to the time horizon $T$, the reconstruction vectors $\vec{v}$ allow for 
(for details see \citep{casdagli1991}). Broadly speaking a low conditional variance, and thus, a low value of
$\langle \sigma(T) \rangle^2$ is achieved for sufficiently unfolded attractors, because in this case noise distortions 
of the true trajectory are not likely to result in mixing states, which are far away from each other in true state
space and consequently preserve the neighbourhood relations on the reconstructed attractor.

\citet{uzal2011} reinterpret Eqs.~\eqref{eq_noise_amplification1}, \eqref{eq_noise_amplification2} and give an 
approximation-recipe for the conditional variance. The authors redefine the mentioned equations as
\begin{equation}\label{eq_uzal_noise_amplification1}
\sigma_{\varepsilon}^2\bigl(\vec{v}_\text{fid}(t)\bigr) = \frac{1}{T_M} \int_0^{T_M} \sigma_{\varepsilon}^2\bigl(T,\vec{v}_\text{fid}(t)\bigr) dT
\end{equation}   
and consider the limit
\begin{equation}\label{eq_uzal_noise_amplification2}
\sigma\bigl(\vec{v}_\text{fid}(t)\bigr) = \lim\limits_{\varepsilon \to 0}\sigma_{\varepsilon}\bigl(\vec{v}_\text{fid}(t)\bigr)
\end{equation}
where $\varepsilon$ is not related to any observational noise level anymore. The $\varepsilon$-ball 
$B_{\varepsilon}\bigl(\vec{v}_\text{fid}(t)\bigr)$ is simply a tool for determining certain neighbourhood relations
of a fiducial point $\vec{v}_\text{fid}(t)$ and their changes when mapped to future states by the
reconstruction function $F$/ $F'$ (Fig.~\ref{fig_embedding_scheme}). It is
possible to approximate the conditional variance in Eq.~\eqref{eq_noise_amplification1} by
\begin{equation}\label{eq_Ek2}
E_k^2(T,\vec{v}_\text{fid}) \equiv \frac{1}{k+1} \sum_{\vec{v}' \in B_k(\vec{v}_\text{fid})} \bigl(\vec{v}'(t+T) - u_k(T,\vec{v}_\text{fid})\bigr)^2
\end{equation}
where $B_k(\vec{v}_\text{fid})$ estimates $B_{\varepsilon}\bigl(\vec{v}_\text{fid}(t)\bigr)$ by the fiducial point itself
and its $k$ nearest neighbours, respecting a \textit{Theiler window} (i.e., avoid temporal correlations in the
neighbour-searching) \cite{theiler1986}. The center of mass with respect to the chosen time horizon $T$ and the
fiducial point $\vec{v}_\text{fid}$ is defined as

\begin{equation}\label{eq_uk}
u_k(T,\vec{v}_\text{fid}) \equiv \frac{1}{k+1} \sum_{\vec{v}' \in B_k(\vec{v}_\text{fid})} \vec{v}'(t+T).
\end{equation}
The size of the $k$-neighbourhood of $\vec{v}_\text{fid}$, $B_k(\vec{v}_\text{fid})$ is estimated as
\begin{equation}\label{eq_epsilonk2}
\epsilon^2_k(\vec{v}_\text{fid}) \equiv \frac{2}{k(k+1)}
\sum\limits_{ \vec{v}', \vec{v}'' \in B_k(\vec{v}_\text{fid})\atop
\vec{v}' \neq \vec{v}''} 
\| \vec{v}' - \vec{v}'' \|^2 ,
\end{equation}
where $\| \cdot \|$ is a norm used for the distance computation. Finally, $E_k^2(T,\vec{v}_\text{fid})$
(Eq.~\eqref{eq_Ek2}) is averaged over a range of $T$'s in $[0, T_M]$ and the noise amplification estimated
from $k$ nearest neighbours is
\begin{equation}\label{eq_sigmak2}
\sigma_k^2(\vec{v}_\text{fid}) \equiv \frac{E_k^2(\vec{v}_\text{fid})}{\epsilon^2_k(\vec{v}_\text{fid})},
\end{equation}
which needs to averaged over all considered fiducial points $N'$ on the reconstructed attractor to obtain
\begin{equation}\label{eq_sigmak2_avrg}
\sigma_k^2 = \sum\limits^{N'}\limits_{i \in \lbrace\vec{v}_\text{fid}\rbrace}\sigma_k^2(\vec{v}_\text{i}).
\end{equation}

Since the reinterpretation of $\varepsilon$ with the related $k$ now acts as a neighbourhood size parameter 
(Eqs.~\ref{eq_uzal_noise_amplification1},\ref{eq_uzal_noise_amplification2},\ref{eq_Ek2},\ref{eq_uk},\ref{eq_epsilonk2},\ref{eq_sigmak2}),
$\sigma_k^2$ can be normalized with respect to the averaged inter-point distance, which depends on the sampling rate and
the scale of the input data \cite{uzal2011}. The normalization factor is
\begin{equation}\label{eq_alphak2}
\alpha_k^2 = \frac{1}{\sum\limits^{N'}\limits_{i \in \lbrace\vec{v}_\text{fid}\rbrace}\epsilon_k^{-2}(\vec{v}_\text{i})}
\end{equation}
with $\epsilon_k^2$ from Eq.~\eqref{eq_epsilonk2}. In this way, the final statistic will be able to 
compare attractor reconstructions stemming from different input data and also serves as an irrelevance measure, because large delays
will result in large $\epsilon_k^2$'s.

Finally, Eqs.~\eqref{eq_sigmak2_avrg},\eqref{eq_alphak2} define the $L$-statistic
\begin{equation}\label{eq_Lstatistic}
L_k = \log_{10}(\alpha_k\sigma_k) ,
\end{equation}
which has a free parameter $k$ and another implicit parameter $T_M$ (Eq.~\eqref{eq_uzal_noise_amplification1}).

\bibliography{embedding_bib}

\begin{thebibliography}{83}
\bibitem [{com({\natexlab{a}})}]{comment_breaking_criterion}%
  \BibitemOpen
  \href@noop {} {\bibfield  {journal} {\bibinfo  {journal} {Implicitly the
  undersampling statistic could be used as a break criterion, when all of the
  considered delays (local maxima in the continuity statistic) are above the
  chosen significance level $\beta$ in the corresponding undersampling
  statistic. But this will not hold for non-chaotic systems and leads to very
  high embedding dimensions in these cases.}\ } ({\natexlab{a}})}\BibitemShut
  {NoStop}%
\end{thebibliography}

\end{document}